\documentclass[journal,letterpaper,twocolumn,final,10pt]{IEEEtran}

\usepackage{amsthm}
\usepackage{amsmath}
\usepackage{amsfonts}
\usepackage{dsfont}
\usepackage{mathrsfs}
\usepackage{pifont}
\usepackage{amssymb}
\usepackage{verbatim}
\usepackage{upgreek}
\usepackage{color}
\usepackage[final]{graphicx}
\usepackage{epstopdf}
\usepackage{algorithmic}
\usepackage{algorithm}
\usepackage{epsfig}
\usepackage{eucal}
\usepackage{cite}
\usepackage{balance}
\usepackage{setspace}
\usepackage{siunitx}
\usepackage{gensymb}

\usepackage{float}
\usepackage{subfig}

\makeatletter
\def\maketag@@@#1{\hbox{\m@th\normalfont\normalsize#1}}
\makeatother

\newcommand{\bGamma}{\mbox{\boldmath{$\Gamma$}}}

\newcommand{\bbm}{\begin{bmatrix}}
\newcommand{\ebm}{\end{bmatrix}}

\newcommand{\bit}{\begin{itemize}}
\newcommand{\eit}{\end{itemize}}

\newcommand{\ben}{\begin{enumerate}}
\newcommand{\een}{\end{enumerate}}

\newcommand{\bdesc}{\begin{description}}
\newcommand{\edesc}{\end{description}}

\newcommand{\bea}{\begin{array}}
\newcommand{\eea}{\end{array}}

\newcommand{\tr}{\mbox{\rm Tr}\, }

\newcommand{\beqa}{\begin{eqnarray}}
\newcommand{\eeqa}{\end{eqnarray}}

\newcommand{\ds}{\displaystyle}

\newcommand{\Comment}[1]{}

\def\C{{\mathds C}}

\def\cC{\mbox{$\CMcal C$}}
\def\cD{\mbox{$\CMcal D$}}

\def\cG{\mbox{$\mathcal G$}}

\def\cN{\mbox{$\CMcal N$}}

\def\cW{\mbox{$\mathcal W$}}

\newcommand{\be}{\begin{equation}}

\newcommand{\ee}{\end{equation}}

\newcommand{\bzero}{{\mbox{\boldmath $0$}}}

\newcommand{\boa}{{\mbox{\boldmath $a$}}}
\newcommand{\bob}{{\mbox{\boldmath $b$}}}

\newcommand{\bg}{{\mbox{\boldmath $g$}}}

\newcommand{\bm}{{\mbox{\boldmath $m$}}}

\newcommand{\bor}{{\mbox{\boldmath $r$}}}
\newcommand{\bw}{{\mbox{\boldmath $w$}}}

\newcommand{\bt}{{\mbox{\boldmath $t$}}}

\newcommand{\bv}{{\mbox{\boldmath $v$}}}
\newcommand{\bx}{{\mbox{\boldmath $x$}}}

\newcommand{\bz}{{\mbox{\boldmath $z$}}}

\newcommand{\bA}{{\mbox{\boldmath $A$}}}
\newcommand{\bB}{{\mbox{\boldmath $B$}}}
\newcommand{\bC}{{\mbox{\boldmath $C$}}}
\newcommand{\bD}{{\mbox{\boldmath $D$}}}

\newcommand{\bG}{{\mbox{\boldmath $G$}}}

\newcommand{\bI}{{\mbox{\boldmath $I$}}}

\newcommand{\bM}{{\mbox{\boldmath $M$}}}

\newcommand{\bP}{{\mbox{\boldmath $P$}}}
\newcommand{\bQ}{{\mbox{\boldmath $Q$}}}
\newcommand{\bR}{{\mbox{\boldmath $R$}}}
\newcommand{\bS}{{\mbox{\boldmath $S$}}}
\newcommand{\bT}{{\mbox{\boldmath $T$}}}
\newcommand{\bU}{{\mbox{\boldmath $U$}}}
\newcommand{\bV}{{\mbox{\boldmath $V$}}}
\newcommand{\bX}{{\mbox{\boldmath $X$}}}
\newcommand{\bY}{{\mbox{\boldmath $Y$}}}
\newcommand{\bW}{{\mbox{\boldmath $W$}}}
\newcommand{\bZ}{{\mbox{\boldmath $Z$}}}

\newcommand{\diag}{\mbox{diag}\, }

\newcommand{\balpha}{{\mbox{\boldmath $\alpha$}}}

\newcommand{\dmax}{\begin{displaystyle}\max\end{displaystyle}}
\newcommand{\dmin}{\begin{displaystyle}\min\end{displaystyle}}

\newcommand{\test}{\mbox{$
\begin{array}{c}
\stackrel{ \stackrel{\textstyle H_1}{\textstyle >} }{
\stackrel{\textstyle <}{\textstyle H_0} }
\end{array}
$}}

\title{RIS-aided Radar Detection Architectures with Application to Low-RCS Targets}

\vspace{0.1cm}

\author{Linjie Yan, Yiding Gao, Fabiola Colone, \IEEEmembership{Senior Member, IEEE}, 
Filippo Costa, \IEEEmembership{Senior Member, IEEE},  
Chengpeng Hao, \IEEEmembership{Senior Member, IEEE}, Giuliano Manara, \IEEEmembership{Life Fellow, IEEE},
and Danilo Orlando$^*$
\thanks{The work of D. Orlando was partially supported by the Italian Ministry of 
Education and Research (MUR) in the framework of the FoReLab project (Departments of Excellence).}
\thanks{F. Colone is with with the Department of Information Engineering, Electronics and 
Telecommunications, Sapienza University of Rome, 00184 Rome, Italy. 
E-mail: {\tt fabiola.colone@uniroma1.it}.}
\thanks{F. Costa, G.Manara, and D. Orlando are with the Department of Information Engineering, Universit\`a di Pisa, 
Via Caruso 16, 56122 Pisa, Italy. 
E-mail: {\tt filippo.costa@unipi.it; giuliano.manara@unipi.it; danilo.orlando@unipi.it}.}
\thanks{Y. Gao, C. Hao, and L. Yan are with the Institute of Acoustics, Chinese Academy 
of Sciences, Beijing 100190, China, and also with the School of Electronic, 
Electrical and Communication Engineering, University of Chinese Academy of
Sciences, Beijing 100049, China. 
E-mail: {\tt gaoyiding@mail.ioa.ac.cn; haochengp@mail.ioa.ac.cn; yanlinjie16@163.com}.}
\thanks{$^{*}$Corresponding author}
}

\begin{document}

\maketitle

\begin{abstract}
In this paper, we address the radar detection of low observable targets
with the assistance of a reconfigurable intelligent surface (RIS). Instead of using
a multistatic radar network as counter-stealth strategy with its 
synchronization, costs, phase coherence, and energy consumption issues,
we exploit a RIS to form a joint monostatic and bistatic configuration that
can intercept the energy backscattered by the target along irrelevant directions
different from the line-of-sight of the radar.
Then, this energy is redirected towards the radar that 
capitalizes all the backscattered
energy to detect the low observable target. To this end, five different
detection architectures are devised that jointly process monostatic
and bistatic echoes and exhibit the constant false alarm rate property
at least with respect to the clutter power. To support the practical implementation,
we also provide a guideline for the design of a RIS that satisfies
the operating requirements of the considered application. The performance
analysis is carried out in comparison with conventional detectors
and shows that the proposed strategy leads to effective solutions
to the detection of low observable targets.
\end{abstract}


\markboth{YAN ET AL.}{RIS-aided Radar Detection Architectures with Application to Low-RCS Targets}

\section{Introduction}
\label{Sec:Introduction}
In the recent years, reconfigurable intelligent surfaces (RISs)
is a technology that has gained a leading role 
in the context of new-generation wireless communications \cite[and references therein]{9311936,9475160}. 
These surfaces can be broadly categorized into passive smart electromagnetic skins or 
reflectarrays, which provide fixed beamforming and fully 
reconfigurable versions capable of dynamic wave control 
\cite{fu2022brief}. Their real-time programmability is enabled 
by semiconductor elements such as PIN and varactor 
diodes \cite{costa2021electromagnetic}, enabling manipulation 
of phase, amplitude, and polarization \cite{sievenpiper2003two}. Advanced 
architectures integrate sensing functions \cite{alamzadeh2021reconfigurable}, 
or active components to 
compensate for double-path loss \cite{wu2022wideband}. This hardware versatility 
positions RIS as a cornerstone for smart programmable radio environments 
in 5G-Advanced and 6G networks.

Although the use of the RIS originates in the field of communication networks, its extension
to sensing systems is making great strides with the advent of integrated sensing and communication (ISAC)
paradigm in urban scenarios \cite{8999605}. As a matter of fact, in this respect, many excellent 
contributions can be found in the open literature focused
on different levels of coexistence, cooperation, or codesign
of the sensing and communication systems. In \cite{10606001},
a design framework for the coexistence of multiple-input multiple-output
(MIMO) radar and communication systems is devised by exploiting the assistance
of a RIS. The codesign metric is the
signal-to-interference-plus-noise ratio (SINR) that is optimized
to enhance the target detection capability of the radar system
and to guarantee the communication quality of service.
Other codesign approaches can be found in \cite{9729741,9264225}
where the RIS is used to mitigate the mutual interference
between the sensing and communication systems. A dual-function radar-communication
system is proposed in \cite{10453326}, where in place of conventional RISs
the so-called hybrid RISs are used to enhance the amount of received energy
backscattered by the target. The fundamental trade-offs for the design
of the radar and communication functions are investigated to
maximize the SINR of the radar while satisfying the
SINR requirements on the communication side.
In the context of 5G/6G, in \cite{11008547}, an orthogonal frequency
division multiplexing based ISAC system is proposed
where a RIS is used by the base station for the detection
of non-cooperative small targets with low radar cross section (RCS).
To this end, the system integrates multiple frames and the
track-before-detect \cite{543865,705887} paradigm is applied to account for target motion.
The use of RIS capable of modulating the polarization is investigated in \cite{11234299}
to obtain an improvement in target classification for a distributed
MIMO ISAC system.

A parallel research line investigates the application
of RIS to radar systems only operating in scenarios that possibly go beyond
those for communications. An important contribution 
is provided in \cite{9454375}, where the authors focus on the system design 
and carry out a theoretical analysis of the main design aspects
including signal bandwidth, carrier frequency, RIS size, and relative 
distance between radar and RIS. The application proposed in \cite{9508883}
considers a radar system operating in a urban scenario where there exist regions 
for which the line-of-sight (LOS) condition is not valid. 
In these environments, the RIS 
comes into play to cover the blind regions. 
The same application
is addressed in \cite{10050582} where optimal phase shifts for the RIS
are computed for each pointing direction belonging to the angular sector
of interest. In \cite{10453357}, a general signal model
for a radar system assisted by RIS is derived by encompassing the LOS and non-LOS
conditions. This model is used to optimize RIS deployment and
for the design of the transmit waveforms, receive filters, 
and RIS phase shifts.
The non-LOS condition is still considered in \cite{10908879}, where multiple RISs are used
to assist a monostatic radar. Specifically, the authors investigate the effects
on the detection performance generated by additional RISs under the non-LOS condition.
A novel radar architecture is developed in \cite{9775578}, where, in addition
to the monostatic echoes, the radar system collects bistatic echoes reflected by the RIS. 
This approach allows for the exploitation of the spatial diversity (on receive)
to enhance the system detection capabilities. Notice that the radar is assumed to illuminate 
the target only and not also the RIS. The potential of RIS 
in the context of MIMO radars
is explored in \cite{9732186} where the devised system processes the returns from 
(up to) two RISs deployed in the region of interest. The RIS phase shifts are computed
in order to maximize the probability of detection ($P_d$) 
for a given probability of false alarm ($P_{fa}$) by means of the
maximum likelihood approach. Other interesting contributions can be found 
in \cite{10948115,10185137,10878502}.

This paper is framed in the second of the two lines of research described above, 
namely the development
of RIS-aided radar systems.
Remarkably, it focuses for the first time (at least to the best
of authors' knowledge) on the design of detection architectures for airborne
targets designed to exhibit a low-RCS. Otherwise stated, the low-RCS is not due
to the target size as assumed in \cite{11008547} but it is a design
objective. As a matter of fact, in the last 
decades, stealth or ``low observable'' technology 
has proven to be one of the most
effective approaches to hide targets from radar 
systems \cite{lowObservable}. 
Especially in the defense context, all new aircraft 
are designed taking into account low observable
principles and techniques, while the project of existing platforms 
is reconsidered to reduce the corresponding radar signature. 
As explained in Section \ref{Sec:Low-RCS}, a design principle of
low-RCS platforms consists in shaping the platform to reduce the backscattered energy
in a specific angular sector (mainly related to the frontal aspect).
More precisely, most of the intercepted energy is redirected towards
{\em irrelevant directions}. Notice that multistatic radar systems represent
an effective means to counter the stealth technology since they use
spatial diversity to intercept the energy reflected towards the aforementioned
irrelevant directions. 

Thus, the main idea behind this paper is that
a monostatic radar can exploit suitably deployed RISs 
to form bistatic configurations capable of intercepting the energy
backscattered by the target along directions different from the LOS.
Unlike \cite{9775578}, here the RIS is placed within the mainbeam
of the radar system. 
This idea has a significant practical value since the use of the RIS
allows for lower costs than a multistatic radar network with its
synchronization, phase coherence, and energy consumption issues.
The RIS is designed and deployed to recover a sufficient level of energy
that can trigger the detection of a low observable target.
To this end, the RIS backscattering
pattern is statically designed to illuminate all the radar cells
under investigation simultaneously with a single, fixed beam.
This is achieved through an adequate, pre-defined tapering of
the surface's unit cells to control both the phase and amplitude
of the reflected wavefront. It is clear that the RIS can be reconfigured
to illuminate different batch of range cells.
Moreover, we investigate the requirements related to the 
geometry of the scenario assuming a search radar with a fan beam, which
also illuminates the RIS,
and provide a formal statement of the
detection problem that accounts for these requirements. Such a problem
appears here for the first time (at least to best of authors' knowledge)
and is solved by applying design criteria based upon the generalized
likelihood ratio test (GLRT) \cite{KayBook} and {\em ad hoc} modifications of it.
To be more definite, since the joint maximization
of the likelihood function with respect to the entire set
of unknown parameters is a difficult task from a mathematical point of view,  
we resort to alternative approaches based upon either
\begin{itemize}
\item the estimate-and-plug paradigm \cite{robey1992cfar,9887896,9888064}
that computes the GLRT assuming that a subset of parameters is known and then
replaces these parameters with suitable estimates;
\item an approximate solution for the GLRT obtained by minimizing
an upper bound to the objective function;
\item an iterative optimization procedure that leads to (at least) local stationary points.
\end{itemize}
Thus, we obtain five different detection architectures and theoretically show 
that they exhibit the constant false 
alarm rate (CFAR) property with respect to the clutter power
in a clutter-dominated environment.
As for the CFAR property with respect to the covariance structure, we
also prove that they are
bounded from above by decision statistics 
functionally invariant to the covariance structure.
Finally, we provide both a technical description of low-RCS target
features and
the guidelines to design a RIS whose task
is the interception of the energy from the irrelevant directions.
The performance analysis, conducted on synthetic data and in comparison
with conventional detectors, clearly shows that the proposed approach
is an effective means to detect low-RCS targets with significant
performance gains over the considered
counterparts.

The remainder of the paper is organized as follows. In the next section, 
the operating requirements are described and the detection problem is formulated 
in terms of a binary hypothesis test. Section \ref{Sec:Architecture_Designs} is devoted to
the design of the detection architectures along with the related estimation procedures.
The analysis of the CFAR behavior is conducted in Section \ref{Sec:CFARproperty}, while
Section \ref{Sec:Low-RCS} is devoted to the design of the RIS according
to the energy requirements for low observable targets. Finally, the performance
assessment is confined to Section \ref{Sec:PerformanceAnalysis} and
some concluding remarks are provided in Section \ref{Sec:Conclusions}.

\subsection{Notation}
In the sequel, vectors and matrices are denoted by boldface lower-case and upper-case letters, respectively.
The $(i,j)$th entry of a matrix $\bA$ is indicated by $\bA(i,j)$.
Symbols $\det(\cdot)$, $\tr(\cdot)$, $(\cdot)^T$, and $(\cdot)^\dag$ denote 
the determinant, trace, transpose,  
and conjugate transpose, respectively.
As to the numerical sets, 
$\C$ is the set of 
complex numbers and $\C^{N\times M}$ is the Euclidean space of $(N\times M)$-dimensional 
complex matrices (or vectors if $M=1$). For two sets $A$ and $B$, the difference set is $A\setminus B$.
Function $\text{rect}(x)$ is the rectangular window and is equal to $1$ if $|x|<1/2$ and to $0$
if $|x|>1/2$.
$\bI$ and $\bzero$ stand for the identity matrix and the null vector or matrix of proper size. 
Given $\boa \in\C^{N\times 1}$, $\diag(\boa)\in\C^{N\times N}$ indicates 
a diagonal matrix whose $i$th diagonal element is the $i$th entry of $\boa$.
If $\bA\in\C^{N\times N}$, $\diag(\bA)\in\C^{N\times N}$ is a 
diagonal matrix whose main diagonal is the same as $\bA$.
Symbol $\|\cdot\|_F$ is the Frobenius norm of the matrix argument.
The acronym PDF stands for Probability Density Function. Finally,
we write $\bx\sim\cC\cN_N(\bm, \bM)$ if $\bx$ is a 
complex circular $N$-dimensional normal vector with mean $\bm$ 
and positive definite covariance matrix $\bM$, whereas
$\bX\sim\cC\cW_N(K,\bA)$ means that matrix $\bX\in\C^{N\times N}$
obeys the complex Wishart distribution with parameters $K$
and $\bA\in\C^{N\times N}$.

\section{Problem Formulation}
\label{Sec:Problem_Formulation}
\begin{figure}[tbp]
\begin{center}
\includegraphics[scale=0.35]{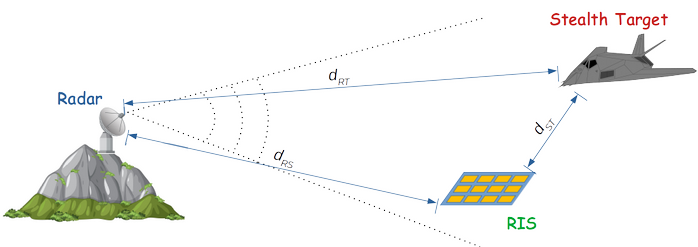}
\caption{Geometry of the operating scenario.}
\label{fig:operatingScenario}
\end{center}
\end{figure}

Let us consider a monostatic radar system that uses a planar array of $N_a$ tiles
to sense the region of interest. The system transmits a coherent burst of $N_p$ pulses 
that are Doppler tolerant \cite{Richards_BP} and the illumination is accomplished by means of
a fan beam.\footnote{The beam shape can be obtained by a suitable tapering of
the spatial channels or by using a subset of sensors along the elevation dimension.} 
A RIS is placed in the region of interest
with the objective of illuminating prospective stealth targets from below
as depicted in Figure \ref{fig:operatingScenario}, where $d_{RS}>0$ is the RIS-radar distance, 
$d_{RT}>0$ is the distance from the radar to the target, and
$d_{ST}>0$ is the distance between the target and the RIS.
The RIS configuration is such that it can manage
the path radar-RIS-target and target-RIS-radar.
The prospective target is shaped in order to exhibit a low radar 
cross section along the flight direction, namely
its surfaces and edges deflect the scattered energy in directions away from
the monostatic radar \cite{knott2004radar,4529}. It is clear that this behavior cannot be forced
for all angles of view and there will always exist aspect
angles at which the target surfaces generate high-energy echoes.
Since many stealth vehicles mainly minimize radar echo in the head-on
direction, a bistatic configuration might allow for the interception
the echoes reflected towards directions that are different from 
that to the electromagnetic source (i.e., the radar system).
Thus, the deployment of a RIS, which is passive and, hence, low observable, 
within the region of interest provides spatial diversity
and leads to a bistatic configuration that can be considered a counter-stealth technology.

\begin{figure}[tbp]
\begin{center}
\includegraphics[scale=0.35]{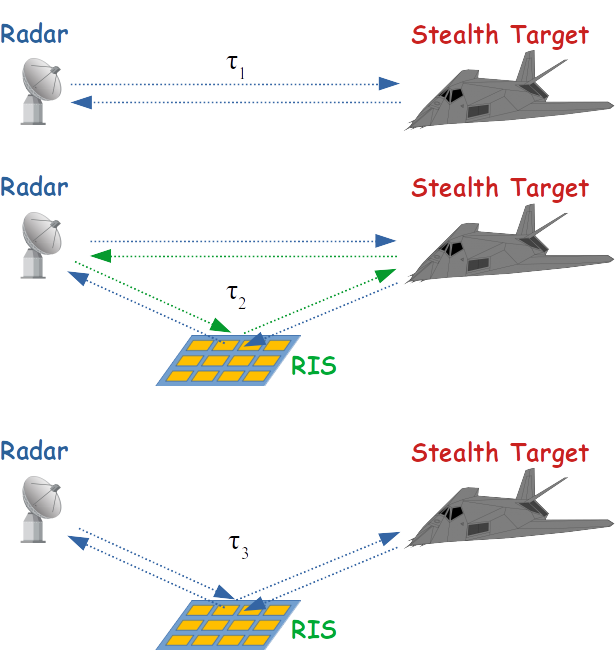}
\caption{Considered paths.}
\label{fig:paths}
\end{center}
\end{figure}

To be more definite, let us consider Figure \ref{fig:paths} where the
most significant routes taken by the received echoes are shown (other
possible paths are neglected). In the figure, the delays $\tau_i$, $i=1,2,3$, associated
with each path are defined as
\be
\begin{cases}
\tau_1 &= \ds\frac{2d_{RT}}{c},
\\
\vspace{-3mm}
\\
\tau_2 &= \ds\frac{d_{RT}+d_{ST}+d_{RS}}{c},
\\
\vspace{-3mm}
\\
\tau_3 &= \ds\frac{2(d_{RS}+d_{ST})}{c},
\end{cases}
\ee
where $c$ is the speed of light. 
For future reference, we denote the paths associated with 
$\tau_1$, $\tau_2$, and $\tau_3$ with radar-target-radar (RTR), 
radar-target-surface-radar (RTSR),
and radar-surface-target-surface-radar (RSTSR), respectively.
Notice that also the path radar-surface-target-radar (RSTR) is associated
with $\tau_2$.
Since the operating scenario is dynamic,
for a target approaching the radar, we can assume that the following inequalities hold
\be
\begin{cases}
\tau_3-\tau_1 &= \ds\frac{2(d_{RS}+d_{ST}-d_{RT})}{c}\ge\frac{4\Delta r}{c}>0,
\\
\vspace{-3mm}
\\
\tau_3-\tau_2 &= \ds\frac{d_{RS}+d_{ST}-d_{RT}}{c}\ge\frac{2\Delta r}{c}>0,
\\
\vspace{-3mm}
\\
\tau_2-\tau_1 &= \ds\frac{d_{RS}+d_{ST}-d_{RT}}{c}\ge\frac{2\Delta r}{c}>0,
\end{cases}
\label{eqn:cond_tau}
\ee
where $\Delta r$ is the radar range resolution.
In fact, in the presence of an airborne target approaching the radar, 
the above conditions can be met for a likely geometry.
For instance, if $d_{RT}=19000$ m, $d_{RS}=20000$ m,
$d_{ST}=1000$ m, and $\Delta r=50$ m, then $\tau_1\approx 126 \ \mu$s, $\tau_2\approx 133 \ \mu$s,
$\tau_3\approx 140 \ \mu$s, and $2 \Delta r/c\approx 0.3 \ \mu$s.
The delay differences are $\tau_3-\tau_1\approx 14 \ \mu$s, $\tau_3-\tau_2\approx 7 \ \mu$s, and
$\tau_2-\tau_1\approx 7 \ \mu$s; all of them are one order of magnitude greater than $2 \Delta r/c$
satisfying the conditions \eqref{eqn:cond_tau}.
As a consequence, if we denote by $k_1$ the index of the range bin containing the echoes
corresponding to $\tau_1$, the reflected signal exhibiting the delay $\tau_2$ occupies
the range cell indexed by $k_2>k_1$, whereas the echoes with delay $\tau_3$ are present
in the range bin indexed by $k_3>k_2$.
Actually, all conditions \eqref{eqn:cond_tau}
are satisfied if
\be
d_{RS}+d_{ST}-d_{RT}\ge 2\Delta r.
\label{eqn:rangeInequality}
\ee
With the above remarks in mind, we jointly process a set of $K_P$ 
consecutive range bins (forming the window under test), where
the first cell is associated with $\tau_1$ and two out of $K_P-1$ range bins
are related to $\tau_2$ and $\tau_3$. The remaining range bins are assumed
free of signal components. Moreover, we assume the availability of 
the conventional set of training samples to estimate the unknown statistical parameters
of the disturbance.\footnote{In what follows, we consider a clutter dominated environment.}
Summarizing, the detection problem of interest can be formulated as 
the following binary hypothesis test
\be
\begin{cases}
H_0: &
\begin{cases}
\bz_k \sim\cC\cN_N(\bzero,\bM), & k\in\Omega_P,
\\
\bor_k \sim\cC\cN_N(\bzero,\bM), & k \in \Omega_S,
\end{cases}
\\
\vspace{-2mm}
\\
H_1: &
\begin{cases}
\bz_1 \sim\cC\cN_N(\alpha_1 \bv_R ,\bM), &
\\
\bz_n \sim\cC\cN_N(\alpha_n \bv_{SR} ,\bM), & n\in\Omega_P, \ n>1,
\\
\bz_m \sim\cC\cN_N(\alpha_m \bv_S ,\bM), & m\in\Omega_P, \ m>n,
\\
\bz_k \sim\cC\cN_N(\bzero,\bM), & k\in\Omega_P\setminus\{1,n,m\},
\\
\bor_k \sim\cC\cN_N(\bzero,\bM), & k\in\Omega_S,
\end{cases}
\end{cases}
\label{eqn:hypothesisTest}
\ee
where the condition $m>n$ makes it possible to account
for effects due to both propagation and fast-time sampling anomalies,
\begin{itemize}
\item $\bz_k\in\C^{N\times 1}$, $k\in\Omega_P=\{1,\ldots,K_P\}$, are the space-time 
vectors (with $N=N_a N_p$ the number of space-time channels) containing
the echo samples from the window under test,\footnote{The incoming signals undergo the conventional
operations of down-conversion, matched filtering, and fast-time sampling.} 
which contains the signals traveling according to
the previously mentioned paths; we denote by 
$\bZ_P=[\bz_1,\ldots,\bz_{K_P}]\in\C^{N\times K_P}$ the overall primary data matrix;
\item $\bor_k\in\C^{N\times 1}$, $k\in\Omega_S=\{1,\ldots,K_S\}$ with 
$K_S\ge N$,\footnote{This condition is required to make
the sample covariance matrix based upon $\bor_1,\ldots,\bor_{K_S}$ 
nonsingular with probability $1$.} are
the training vectors
assumed free of useful signal components and statistically
independent of $\bz_1,\ldots,\bz_{K_P}$;
we denote by $\bR=[\bor_1,\ldots,\bor_{K_S}]\in\C^{N\times K_S}$ the overall
secondary data matrix;
\item $\bv_R\in\C^{N\times 1}$ is the known space-time target steering vector as seen by the 
radar and, hence, is a function of target azimuth angle, elevation angle, and monostatic
Doppler frequency (computed with respect to the radar);
\item $\bv_{SR}=\bv_{S,1}+\bv_{R,1}$ is the composition between the
steering vectors associated with each path corresponding to $\tau_2$ (see
Figure \ref{fig:paths}); specifically, $\bv_{S,1}\in\C^{N\times 1}$ is the steering 
vector associated with the path RTSR
and $\bv_{R,1}\in\C^{N\times 1}$ the steering vector corresponding to the path RSTR, both
characterized by the same bistatic Doppler frequency but different angles of arrival;
\item $\bv_S\in\C^{N\times 1}$ is the known space-time target signature 
of the RIS with respect to the radar when it
reflects target echoes (in this case azimuth and elevation angles are those of the RIS
while Doppler frequency depends on the target radial velocity with respect to the RIS);
\item $\alpha_1\in\C$, $\alpha_n\in\C$, and $\alpha_m\in\C$ are the unknown target responses, which
also include channel and system effects, corresponding to 
the paths RTR, RTSR (RSTR), and RSTSR, respectively;
\item $\bM\in\C^{N\times N}$ is the unknown positive definite 
covariance matrix of the disturbance.
\end{itemize}
The subsequent derivations require 
the expressions of the joint PDFs of $\bZ_P$
and $\bR$ 
under each hypothesis. Specifically, denoting by $K=K_P+K_S$ the total number of vectors, 
the PDFs under $H_0$ and $H_1$ can be written as
\be
f(\bZ_P,\bR; \bM, H_0)=\frac{\exp\{-\tr[\bM^{-1}(\bS_P+\bS_S)]\}}{\pi^{KN} \det^K(\bM)}
\label{eqn:PDF_H0}
\ee
and
\begin{multline}
f(\bZ_P,\bR;\balpha,n,m,\bM, H_1)=\frac{1}{\pi^{KN} \det^K(\bM)}
\\
\exp\Bigg\{  
-\tr\Bigg[\bM^{-1}
\Bigg(
(\bZ_{1,n,m}-\bV\bA)
\\
\times (\bZ_{1,n,m}-\bV\bA)^\dag
+\bS_{n,m}
\Bigg)
\Bigg]
\Bigg\},
\label{eqn:PDF_H1}
\end{multline}
respectively, where $\bS_P=\bZ_P\bZ_P^\dag$, $\bS_S=\bR\bR^\dag$, 
\be
\bS_{n,m}=\bS_S+\sum_{k\in\Omega_P\setminus\{1,n,m\}}\bz_k\bz_k^\dag,
\label{eqn:Snm}
\ee
$\bZ_{1,n,m}=[\bz_1,\bz_n,\bz_m]\in\C^{N\times 3}$,
$\balpha=[\alpha_1,\alpha_n,\alpha_m]^T\in\C^{3\times 1}$,
$\bA=\diag(\balpha)$, 
and\footnote{For
simplicity, we have omitted the dependence of $\balpha$ on $(n,m)$.} 
$\bV=[\bv_R,\bv_{SR},\bv_S]\in\C^{N\times 3}$. Notice that
in the case of space-time processing and assuming that
the bistatic Doppler is different from the monostatic Doppler,
$\bV$ is full-column rank.\footnote{In the ensuing derivations,
this assumption is required to come up with closed form estimates. However,
as we will see, the final decision statistics are independent of this assumption 
and, hence, they can be also used for space processing only.}

In the next section, we devise detection architectures for problem \eqref{eqn:hypothesisTest}
by exploiting the GLRT design paradigm and suitable modifications of it that allow us
to overcome possible intractable mathematics.

\section{Architecture Designs}
\label{Sec:Architecture_Designs}
The general formulation of the GLRT for problem \eqref{eqn:hypothesisTest} is
\be
\frac{\dmax_{n,m\in\Omega_P \atop 
n>1,m>n}\dmax_{\bM}\dmax_{\balpha}f(\bZ_P,\bR;\balpha,n,m,\bM, H_1)}
{\dmax_{\bM}f(\bZ_P,\bR;\bM, H_0)}
\test\eta,
\label{eqn:GLRT_generalForm}
\ee
where $\eta$ is the detection threshold whose value depends on the
$P_{fa}$ specified by system 
requirements.\footnote{Hereafter, we denote by $\eta$ the generic detection
threshold for each architecture.}
The maximization at the denominator is a well-known problem that has been
addressed in a plethora of available contributions
(see for instance \cite{kelly1986adaptive,GLRT-based,MultichannelAdaptiveSignal,
BOR-Morgan,6757035,7762200,1396421,9887896} and references therein). Thus, for brevity,
we recall here that the expression of the compressed
likelihood function under $H_0$ is\footnote{The term compressed means
that the likelihood function has been maximized with respect to the unknown parameters.} 
\be
f(\bZ_P,\bR;\widehat{\bM}, H_0)=\left[\frac{K}{\pi e \det^{1/N}(\bS_P+\bS_S)}
\right]^{KN}
\label{eqn:compressed_H0}
\ee
and it will be used in the next subsections when required.

On the other hand, the optimization of the likelihood function 
under $H_1$ (i.e., the numerator of \eqref{eqn:GLRT_generalForm})
is involved from a mathematical point of view and, hence, we resort to suboptimal
modifications of the maximum likelihood approach to come up with suitable
estimates of the unknown parameters. In the next subsections, we describe
these modifications that lead to different decision schemes.

\subsection{Estimate-and-Plug Solutions}
The approach used in this subsection (also known as 
two-step GLRT \cite{robey1992cfar}) consists
in deriving the GLRT when a subset of parameters are known. As a consequence,
the resulting decision statistic is a function of 
such parameters. To make it adaptive,
the remaining parameters are replaced by suitable estimates.

Let us start by considering the following steps
\begin{enumerate}
\item compute the GLR for known $\bM$, that is
\be
\frac{\dmax_{n,m\in\Omega_P \atop 
n>1,m>n}\dmax_{\balpha}f(\bZ_P,\bR;\balpha,n,m,\bM, H_1)}
{f(\bZ_P,\bR;\bM, H_0)};
\label{eqn:GLR_MKnown}
\ee
\item replace $\bM$ with a suitable 
estimate obtained from the 
available data (a point better explained below).
\end{enumerate}
Thus, given $n>1$ and $m>n$, the maximization with respect to $\balpha$ is tantamount to
\be
\dmin_{\balpha}\tr\left\{\bM^{-1}
\left[
(\bZ_{1,n,m}-\bV\bA)(\bZ_{1,n,m}-\bV\bA)^\dag
\right]\right\},
\ee
where the objective function is convex. Thus, it is not difficult to show that
setting to zero the gradient with respect to $\balpha$, we come up with
the following minimizers
\begin{align}
\widehat{\alpha}_1(\bM) &= \frac{\bv_R^\dag\bM^{-1}\bz_1}{\bv_R^\dag\bM^{-1}\bv_R},
\label{eqn:estimate_a1}
\\
\widehat{\alpha}_n(\bM) &= \frac{\bv_{SR}^\dag\bM^{-1}\bz_n}{\bv_{SR}^\dag\bM^{-1}\bv_{SR}},
\label{eqn:estimate_an}
\\
\widehat{\alpha}_m(\bM) &= \frac{\bv_S^\dag\bM^{-1}\bz_m}{\bv_S^\dag\bM^{-1}\bv_S}.
\label{eqn:estimate_am}
\end{align}
Replacing these estimates in \eqref{eqn:GLR_MKnown} and taking the natural logarithm, 
we obtain the following statistically equivalent test
\begin{multline}
\dmax_{n,m\in\Omega_P \atop 
n>1,m>n}\left\{
\frac{|\bv_R^\dag\bM^{-1}\bz_1|^2}{\bv_R^\dag\bM^{-1}\bv_R}
+\frac{|\bv_{SR}^\dag\bM^{-1}\bz_n|^2}{\bv_{SR}^\dag\bM^{-1}\bv_{SR}}\right.
\\
\left.+\frac{|\bv_{S}^\dag\bM^{-1}\bz_m|^2}{\bv_{S}^\dag\bM^{-1}\bv_{S}}
\right\}\test \eta.
\label{eqn:EPGLRT_NA}
\end{multline}
The {\em plug} step is accomplished by replacing $\bM$ with 
$\bS_S$ (that is $K$ times the maximum likelihood estimate of $\bM$ over $\bR$)
or $\bS_{n,m}$ (see \eqref{eqn:Snm}).
Finally, the maximization with respect to $(n,m)$ is numerically conducted 
over a grid of points. In what follows, we refer to \eqref{eqn:EPGLRT_NA}
coupled with $\bS_S$ as estimate-and-plug GLRT for known $\bM$ solution $1$ (EP-GLRT-KM-1),
while \eqref{eqn:EPGLRT_NA} coupled with $\bS_{n,m}$ is called 
estimate-and-plug GLRT for known $\bM$ solution $2$ (EP-GLRT-KM-2).

As an alternative strategy, instead of considering $\bM$ known, we can assume that
$\balpha$ is given. Thus, the modified procedure consists of the following steps
\begin{enumerate}
\item compute the GLR for known $\balpha$, that is
\be
\frac{\dmax_{n,m\in\Omega_P \atop 
n>1,m>n}\dmax_{\bM}f(\bZ_P,\bR;\balpha,n,m,\bM, H_1)}
{\dmax_{\bM}f(\bZ_P,\bR;\bM, H_0)};
\label{eqn:GLR_aKnown}
\ee
\item replace the entries of $\balpha$ with suitable estimates.
\end{enumerate}
Thus, given $n>1$ and $m>n$, the maximization with respect to $\bM$ under $H_1$ leads 
to\footnote{We omit the derivation as it is similar to that under $H_0$.}
\begin{multline}
f(\bZ_P,\bR;\balpha,n,m,\widehat{\bM}, H_1)
\\
=\left[
\frac{K}{\pi e \det^{1/N}\left(\bY_{1,n,m}+\bS_{n,m}\right)}
\right]^{KN},
\label{eqn:compressed_aKnown_H1}
\end{multline}
where $\bY_{1,n,m}=(\bZ_{1,n,m}-\bV\bA)(\bZ_{1,n,m}-\bV\bA)^\dag$.
Exploiting \eqref{eqn:compressed_aKnown_H1} and \eqref{eqn:compressed_H0}, the GLRT for known $\balpha$
can be equivalently recast as
\be
\dmax_{n,m\in\Omega_P \atop 
n>1,m>n}\frac{\det(\bS_P+\bS_S)}
{\det[\bS_{n,m} + \bY_{1,n,m}]}
\test\eta
\label{eqn:EP-GLRT-KA}
\ee
and the adaptive GLRT is obtained by replacing $\balpha$ with
\be
\widehat{\balpha}=\begin{bmatrix}
\widehat{\alpha}_1(\bS_S)
\\
\widehat{\alpha}_n(\bS_S) 
\\
\widehat{\alpha}_m(\bS_S)
\end{bmatrix}.
\ee
In what follows, we refer to this detector as 
estimate-and-plug GLRT for known $\balpha$ (EP-GLRT-KA).

\subsection{GLRT based on Cyclic Optimization}
\label{subsec:C-GLRT}
The estimation procedure of the unknown parameters under $H_1$
described in this subsection is grounded on cyclic optimization strategies 
that lead to a nondecreasing sequence of likelihood values \cite{Stoica_alternating}.
Specifically, let us focus again on 
the numerator of \eqref{eqn:GLRT_generalForm} and solve
the maximization with respect to $\bM$. 
The partially-compressed likelihood function is further optimized by
repeating an iteration formed by the following steps 
until a stopping criterion is not satisfied
\begin{enumerate}
\item assume that $\alpha_n$ and $\alpha_m$ are known and equal to the estimates
obtained in the previous iteration, then maximize the partially-compressed likelihood
with respect to $\alpha_1$;
\item let $\alpha_1$ be equal to the estimate obtained in the previous step and
assume again that $\alpha_m$ is equal to the value in the previous iteration, then
maximize the partially-compressed likelihood with respect to $\alpha_n$;
\item assume that $\alpha_1$ and $\alpha_n$ are equal 
to the estimates obtained in the steps
$1$ and $2$, respectively, and maximize the partially-compressed 
likelihood with respect to $\alpha_m$.
\end{enumerate}
Denoting by $g(\alpha_1,\alpha_n,\alpha_m)=f(\bZ_P,\bR;\balpha,n,m,\widehat{\bM}, H_1)$
the partially-compressed likelihood function (with respect to
$\bM$), iterating the above steps leads
to the following inequality chain
\begin{multline}
g(\alpha^{(h-1)}_1,\widehat{\alpha}^{(h-1)}_n,\widehat{\alpha}^{(h-1)}_m)
\leq
g(\widehat{\alpha}^{(h)}_1,\widehat{\alpha}^{(h-1)}_n,\widehat{\alpha}^{(h-1)}_m)
\\
\leq
g(\widehat{\alpha}^{(h)}_1,\widehat{\alpha}^{(h)}_n,\widehat{\alpha}^{(h-1)}_m)
\leq
g(\widehat{\alpha}^{(h)}_1,\widehat{\alpha}^{(h)}_n,\widehat{\alpha}^{(h)}_m)
\\
\leq
g(\widehat{\alpha}^{(h+1)}_1,\widehat{\alpha}^{(h)}_n,\widehat{\alpha}^{(h)}_m)
\leq \ldots,
\end{multline}
where $h$ is the iteration index, $\widehat{\alpha}^{(h-1)}_i$, $i\in\{n,m\}$,
are the estimation updates in the $(h-1)$th iteration,
$\widehat{\alpha}^{(h)}_1$ is obtained
through step $1$, $\widehat{\alpha}^{(h)}_n$ is obtained by means of
step $2$, and $\widehat{\alpha}^{(h)}_m$ comes from step $3$.
As for the stopping criterion, let us define
\begin{multline}
\cG(\widehat{\alpha}^{(h)}_1,\widehat{\alpha}^{(h)}_n,\widehat{\alpha}^{(h)}_m)
\\
=\frac{g(\widehat{\alpha}^{(h)}_1,\widehat{\alpha}^{(h)}_n,\widehat{\alpha}^{(h)}_m)
-g(\widehat{\alpha}^{(h-1)}_1,\widehat{\alpha}^{(h-1)}_n,\widehat{\alpha}^{(h-1)}_m)}
{g(\widehat{\alpha}^{(h-1)}_1,\widehat{\alpha}^{(h-1)}_n,\widehat{\alpha}^{(h-1)}_m)},
\end{multline}
then, a possible choice is
\be
0\le\cG(\widehat{\alpha}^{(h)}_1,\widehat{\alpha}^{(h)}_n,\widehat{\alpha}^{(h)}_m)<\epsilon,
\label{eqn:stoppingCriterion}
\ee
or $h\leq h_{\max}$, where $h_{\max}$ is the maximum allowed number 
of iterations and $\epsilon>0$. 
Parameters $h_{\max}$ and $\epsilon$ can be set according to 
computational and performance requirements.

Therefore, to find the expression of the estimation updates, 
we apply (for computational convenience) 
the logarithm to the partially-compressed likelihood function
and consider 
\begin{align}
&\log g(\alpha_1,\widehat{\alpha}^{(h-1)}_n,\widehat{\alpha}^{(h-1)}_m) = C 
-K\log\det\Big[\bS_{n,m}
\nonumber
\\
&+\left(\bz_1-\alpha_1\bv_R\right)\left(\bz_1-\alpha_1\bv_R\right)^\dag
+\left(\bZ_{n,m}-\bV_{SRS}\bA_{n,m}^{(h-1)}\right)
\nonumber
\\
&\times
\left(\bZ_{n,m}-\bV_{SRS}\bA_{n,m}^{(h-1)}\right)^\dag\Big],
\label{eqn:compressedLL_unKnown_alpha_1}
\end{align}
where $C=KN\log[K/(\pi e)]$, $\bZ_{n,m}=[\bz_n,\bz_m]$, $\bV_{SRS}=[\bv_{SR},\bv_S]$,
and $\bA_{n,m}^{(h-1)}=\diag([\widehat{\alpha}^{(h-1)}_n,\widehat{\alpha}^{(h-1)}_m]^T)$. 
Now, let us define 
\begin{multline}
\bC_{n,m}^{(h-1)}=\bS_{n,m}+\left(\bZ_{n,m}-\bV_{SRS}\bA_{n,m}^{(h-1)}\right)
\\
\times
\left(\bZ_{n,m}-\bV_{SRS}\bA_{n,m}^{(h-1)}\right)^\dag,
\end{multline}
then \eqref{eqn:compressedLL_unKnown_alpha_1} can be recast as
\begin{multline}
C-K\log\det\left[\bC_{n,m}^{(h-1)}\right]-K
\\
\times\log\left[1+\left(\bz_1-\alpha_1\bv_R\right)^\dag
(\bC_{n,m}^{(h-1)})^{-1}
\left(\bz_1-\alpha_1\bv_R\right)\right].
\end{multline}
Maximizing the above function with respect to $\alpha_1$ is tantamount to minimizing
the argument of the logarithm, namely
\be
\min_{\alpha_1} \left[\left(\bz_1-\alpha_1\bv_R\right)^\dag
(\bC_{n,m}^{(h-1)})^{-1}
\left(\bz_1-\alpha_1\bv_R\right)\right]
\ee
that is a well-known problem leading to
\be
\widehat{\alpha}_1^{(h)}=\frac{\bv_R^\dag(\bC_{n,m}^{(h-1)})^{-1}\bz_{1}}
{\bv_R^\dag(\bC_{n,m}^{(h-1)})^{-1}\bv_R}.
\ee
Now, according to step $2$, we consider
\begin{align}
&\log g(\widehat{\alpha}_1^{(h)},\alpha_n,\widehat{\alpha}^{(h-1)}_m) = C 
-K\log\det\Big[\bS_{n,m}
\nonumber
\\
&+\left(\bz_n-\alpha_n\bv_{SR}\right)\left(\bz_n-\alpha_n\bv_{SR}\right)^\dag
+\left(\bZ_{1,m}-\bV_{RS}\bA_{1,m}^{(h-1)}\right)
\nonumber
\\
&\times
\left(\bZ_{1,m}-\bV_{RS}\bA_{1,m}^{(h-1)}\right)^\dag\Big],
\label{eqn:compressedLL_unKnown_alpha_n}
\end{align}
where $\bV_{RS}=[\bv_R,\bv_S]$, 
$\bA_{1,m}^{(h-1)}=\diag([\widehat{\alpha}_1^{(h)},\widehat{\alpha}^{(h-1)}_m]^T)$,
and $\bZ_{1,m}=[\bz_1,\bz_m]$. Following the same line of reasoning as for $\alpha_1$,
we come up with the following update for the estimate of $\alpha_n$
\be
\widehat{\alpha}_n^{(h)}=\frac{\bv_{SR}^\dag(\bC_{1,m}^{(h-1)})^{-1}\bz_{n}}
{\bv_{SR}^\dag(\bC_{1,m}^{(h-1)})^{-1}\bv_{SR}},
\ee
where
\begin{multline}
\bC_{1,m}^{(h-1)}=\bS_{n,m}+\left(\bZ_{1,m}-\bV_{RS}\bA_{1,m}^{(h-1)}\right)
\\
\times
\left(\bZ_{1,m}-\bV_{RS}\bA_{1,m}^{(h-1)}\right)^\dag.
\end{multline}
Finally, we perform step $3$ of the $h$th iteration by considering
\begin{align}
&\log g(\widehat{\alpha}_1^{(h)},\widehat{\alpha}_n^{(h)},\alpha_m) = C 
-K\log\det\Big[\bS_{n,m}
\nonumber
\\
&+\left(\bz_m-\alpha_m\bv_{S}\right)\left(\bz_m-\alpha_m\bv_{S}\right)^\dag
+\left(\bZ_{1,n}-\bV_{RSR}\bA_{1,n}^{(h)}\right)
\nonumber
\\
&\times
\left(\bZ_{1,n}-\bV_{RSR}\bA_{1,n}^{(h)}\right)^\dag\Big],
\label{eqn:compressedLL_unKnown_alpha_m}
\end{align}
where $\bZ_{1,n}=[\bz_1,\bz_n]$, 
$\bA_{1,n}^{(h)}=\diag([\widehat{\alpha}_1^{(h)},\widehat{\alpha}_n^{(h)}]^T)$,
and $\bV_{RSR}=[\bv_R,\bv_{SR}]$. It is not difficult to show that 
the optimization of the above function
with respect to $\alpha_m$ leads to
\be
\widehat{\alpha}_m^{(h)}=\frac{\bv_{S}^\dag(\bC_{1,n}^{(h)})^{-1}\bz_{n}}
{\bv_{S}^\dag(\bC_{1,n}^{(h)})^{-1}\bv_{S}},
\ee
where
\begin{multline}
\bC_{1,n}^{(h)}=\bS_{n,m}+\left(\bZ_{1,n}-\bV_{RSR}\bA_{1,n}^{(h)}\right)
\\
\times
\left(\bZ_{1,n}-\bV_{RSR}\bA_{1,n}^{(h)}\right)^\dag.
\end{multline}
Finally, denoting by $\bar{h}$ the total number of iterations determined
by the stopping criterion (see \eqref{eqn:stoppingCriterion}), 
the cyclic GLRT (C-GLRT) is statistically equivalent to
\be
\dmax_{n,m\in\Omega_P \atop {n>1, m>n}}\frac{\det(\bS_P+\bS_S)}
{\det\left[\bS_{n,m} + \bar{\bY}_{1,n,m}
\bar{\bY}_{1,n,m}^\dag\right]}
\test\eta,
\label{eqn:C-GLRT_final}
\ee
where $\bar{\bY}_{1,n,m}=\bZ_{1,n,m}-\bV\widehat{\bA}$ and
$\widehat{\bA}=\diag([\widehat{\alpha}^{(\bar{h})}_1,
\widehat{\alpha}^{(\bar{h})}_n,\widehat{\alpha}^{(\bar{h})}_m]^T)$.

\subsection{Approximate GLRT}
The final approach consists in minimizing an upper bound to the objective function
instead of the objective function itself. To be more definite, let us consider the following problem
\begin{align}
&\dmax_{n,m\in\Omega_P \atop 
n>1,m>n}\dmax_{\balpha} f(\bZ_P,\bR;\balpha,n,m,\widehat{\bM}, H_1)
\\
&=\dmax_{n,m\in\Omega_P \atop 
n>1,m>n}\dmax_{\balpha}\left\{\left[
\frac{K}{\pi e}
\right]^{KN} \left[ \det\left(\bS_{n,m}+ \bY_{1,n,m}\right)\right]^{-K}
\right\}
\end{align}
and notice that the maximization with respect to $\balpha$ requires
to solve
\begin{align}
&\dmin_{\balpha} \det\left[
\bS_{n,m}+ \bY_{1,n,m}
\right]
\nonumber
\\
&\Rightarrow \dmin_{\balpha} \det\left[
\bI + (\bZ_{1,n,m}-\bV\bA)^\dag\bS_{n,m}^{-1}(\bZ_{1,n,m}-\bV\bA)
\right]
\nonumber
\\
&\Rightarrow \dmin_{\balpha} \det\left[
\bI + (\bX_{1,n,m}-\bV_S\bA)^\dag(\bX_{1,n,m}-\bV_S\bA)
\right]
\nonumber
\\
&\Rightarrow \dmin_{\balpha} \det\left[\bI+\bX_{1,n,m}^\dag\bX_{1,n,m}\right.
\nonumber
\\
&
+\left(
(\bV_S^\dag\bV_S)\bA-\bV_S^\dag\bX_{1,n,m}\right)^\dag (\bV_S^\dag\bV_S)^{-1}
\nonumber
\\
&\left.\times\left(\bV_S^\dag\bV_S)\bA-\bV_S^\dag\bX_{1,n,m}\right)
-\bX_{1,n,m}^\dag\bP_{V_S}\bX_{1,n,m}\right]
\nonumber
\\
&\Rightarrow \dmin_{\balpha} \det\left[\bI+\bX_{1,n,m}^\dag\bP^\perp_{V_S}\bX_{1,n,m}
\right.
\nonumber
\\
&\left.+\left(\bB\bA-\bG_{1,n,m} \right)^\dag\left(\bB\bA-\bG_{1,n,m} \right)\right],
\label{eqn:minimizationProblemA}
\end{align}
where\footnote{Notice that it is understood that
$\bV$ is full-column rank.}
$\bX_{1,n,m}=\bS_{n,m}^{-1/2}\bZ_{1,n,m}$, 
$\bV_S=\bS_{n,m}^{-1/2}\bV$, 
$\bP^\perp_{V_S}=\bI-\bV_S(\bV_S^\dag\bV_S)^{-1}\bV_S^\dag$, 
$\bB=(\bV_S^\dag\bV_S)^{1/2}$,
and $\bG_{1,n,m}=(\bV_S^\dag\bV_S)^{-1/2}\bV_S^\dag\bX_{1,n,m}$.
If $\bA$ was a generic matrix, the minimizer would be
\be
\tilde{\bA}=\bB^{-1}\bG_{1,n,m}=(\bV_S^\dag\bV_S)^{-1}\bV_S^\dag\bX_{1,n,m}.
\label{eqn:solutionGeneric}
\ee
However, due to the structure of $\bA$, the above estimate
does not minimize the objective function in \eqref{eqn:minimizationProblemA}.
Nevertheless, we resort to another approach to somehow solve \eqref{eqn:minimizationProblemA}.
Specifically, since the logarithm is a monotone increasing function, the minimizer
of the last problem is the same as that of the following problem
\be
\dmin_{\balpha} \log\det\big[
\bI+\bX_{1,n,m}^\dag\bP^\perp_{V_S}\bX_{1,n,m} +\bD^{n,m}_\alpha
\big],
\ee
where $\bD^{n,m}_\alpha=\left(\bB\bA-\bG_{1,n,m} \right)^\dag\left(\bB\bA-\bG_{1,n,m} \right)$.
In addition, instead of minimizing the last objective function, we observe that
for any positive definite matrix $\bX$, the following inequality holds \cite{lutkepohl1997handbook}
\be
\log\det(\bX)\leq \tr(\bX)-N
\ee
and, hence, 
\begin{multline}
\dmin_{\balpha} \log\det\Big[
\bI+\bX_{1,n,m}^\dag\bP^\perp_{V_S}\bX_{1,n,m} 
+\bD^{n,m}_\alpha\Big]
\\
\le
\dmin_{\balpha} \Big\{\tr\Big[
\bI+\bX_{1,n,m}^\dag\bP^\perp_{V_S}\bX_{1,n,m} +\bD^{n,m}_\alpha
\Big]-N\Big\}.
\end{multline}
Therefore, we minimize the above upper bound with respect to $\balpha$, namely
\begin{align}
&\dmin_{\balpha} \tr\left[\left(\bB\bA-\bG_{1,n,m} \right)^\dag\left(\bB\bA-\bG_{1,n,m} \right)\right]
\\
&\Rightarrow  \dmin_{\balpha}\left\|
\bB\bA-\bG_{1,n,m}
\right\|_F^2.
\end{align}
Thus, we find an approximate solution by minimizing the Frobenius norm 
of $\bB\bA-\bG_{1,n,m}$, that is
\begin{multline}
\dmin_{\balpha}\Big[
(\alpha_1\bob_1-\bg_1)^\dag(\alpha_1\bob_1-\bg_1)+(\alpha_n\bob_2-\bg_2)^\dag
\\
\times (\alpha_n\bob_2-\bg_2)+(\alpha_m\bob_3-\bg_3)^\dag(\alpha_m\bob_3-\bg_3)\Big],
\end{multline}
where $\bB=[\bob_1,\bob_2,\bob_3]$ 
and $\bG_{1,n,m}=[\bg_1,\bg_2,\bg_3]$ (again, in order not to burden
the notation we have omitted the dependence on $n$ and $m$).
Setting to zero the gradient of the above objective function, we come up with
\be
\bar{\alpha}_1=\frac{\bob_1^\dag\bg_1}{\bob_1^\dag\bob_1}, \ 
\bar{\alpha}_n=\frac{\bob_2^\dag\bg_2}{\bob_2^\dag\bob_2}, \ 
\bar{\alpha}_m=\frac{\bob_3^\dag\bg_3}{\bob_3^\dag\bob_3}.
\ee
Moreover, let us observe that
\begin{align}
&\left[\diag([\bob_1^\dag\bob_1,\bob_2^\dag\bob_2,\bob_3^\dag\bob_3]^T)\right]^{-1}\bB^\dag\bG_{1,n,m}
\nonumber
\end{align}
\begin{align}
&=
\begin{bmatrix}
\ds
\frac{\bob_1^\dag\bg_1}{\bob_1^\dag\bob_1} & \ds\frac{\bob_1^\dag\bg_2}{\bob_1^\dag\bob_1} & 
\ds\frac{\bob_1^\dag\bg_3}{\bob_1^\dag\bob_1}
\\
\ds
\frac{\bob_2^\dag\bg_1}{\bob_2^\dag\bob_2} & \ds\frac{\bob_2^\dag\bg_2}{\bob_2^\dag\bob_2} & 
\ds\frac{\bob_2^\dag\bg_3}{\bob_2^\dag\bob_2}
\\
\ds
\frac{\bob_3^\dag\bg_1}{\bob_3^\dag\bob_3} & \ds\frac{\bob_3^\dag\bg_2}{\bob_3^\dag\bob_3} & 
\ds\frac{\bob_3^\dag\bg_3}{\bob_3^\dag\bob_3}
\end{bmatrix}
\nonumber
\\
&=\left[\diag(\bV_S^\dag\bV_S)\right]^{-1}\bV_S^\dag\bX_{1,n,m}
\end{align}
and, hence, we obtain
\begin{align}
\bar{\alpha}_1 &= \frac{\bv_R^\dag\bS_{n,m}^{-1}\bz_1}{\bv_R^\dag\bS_{n,m}^{-1}\bv_R}=\widehat{\alpha}_1(\bS_{n,m}),
\label{eqn:estimate_a1_appr}
\\
\bar{\alpha}_n &= \frac{\bv_{SR}^\dag\bS_{n,m}^{-1}\bz_n}{\bv_{SR}^\dag\bS_{n,m}^{-1}\bv_{SR}}
=\widehat{\alpha}_n(\bS_{n,m}),
\label{eqn:estimate_an_appr}
\\
\bar{\alpha}_m &= \frac{\bv_S^\dag\bS_{n,m}^{-1}\bz_m}{\bv_S^\dag\bS_{n,m}^{-1}\bv_S}=\widehat{\alpha}_m(\bS_{n,m}).
\label{eqn:estimate_am_appr}
\end{align}
Two remarks are now in order. First, the constraint on the rank of $\bV$ is no longer
required. Second, observe that these solutions cannot be obtained by assigning to
$\alpha_1$, $\alpha_n$, and $\alpha_m$ the entries $\tilde{\bA}(1,1)$, $\tilde{\bA}(2,2)$, 
and $\tilde{\bA}(3,3)$ given by
\eqref{eqn:solutionGeneric}, respectively. As a matter of fact, in 
\eqref{eqn:solutionGeneric}, the term $\bV_S^\dag\bX_{1,n,m}$ is multiplied 
(from the left) by the inverse
of $\bV_S^\dag\bV_S$ that is different from a diagonal matrix whose nonzero elements
are the inverse of the terms in the principal diagonal of $\bV_S^\dag\bV_S$.

The final expression of the decision rule is given by
\be
\dmax_{n,m\in\Omega_P \atop 
n>1,m>n}\frac{\det(\bS_P+\bS_S)}
{\det\left[
\bS_{n,m} + \bQ_{n,m}
\right]}\test \eta,
\label{eqn:A-GLRT_final}
\ee
where $\bQ_{n,m}=(\bZ_{1,n,m}-\bV\bar{\bA})(\bZ_{1,n,m}-\bV\bar{\bA})^\dag$ and
$\bar{\bA}=\diag([\bar{\alpha}_1,\bar{\alpha}_n,\bar{\alpha}_m]^T)$.
The above decision scheme is referred to in the following as approximate GLRT (A-GLRT).

\section{On the CFAR Property}
In this section, we investigate the CFAR behavior of the proposed decision schemes with
respect to the unknown covariance matrix of the disturbance. To this end, 
we distinguish between the disturbance power and correlation by writing the
covariance matrix as the
product of a scaling factor $\gamma^2>0$ (representing the disturbance power) and
a covariance structure $\bGamma\in\C^{N\times N}$ (that depends on the disturbance correlation shape), i.e.,
$\bM=\gamma^2\bGamma$.\footnote{In a clutter-dominated environment,
the disturbance power coincides with the clutter power.}

Let us start from the EP-GLRT-KM-1 and write
$\bz_1=\gamma\bw_1$, $\bz_i=\gamma\bw_i$, $i=n,m$, and $\bor_k=\gamma\bt_k$, $k\in\Omega_S$.
Under $H_0$, for $i=n,m$ and $k\in\Omega_S$,  $\bw_1,\bw_i,\bt_k\sim\cC\cN_N(\bzero,\bGamma)$.
Then, the decision statistics can be written as
\begin{multline}
\dmax_{n,m\in\Omega_P \atop 
n>1,m>n}\left\{
\frac{|\bv_R^\dag\bT^{-1}\bw_1|^2}{\bv_R^\dag\bT^{-1}\bv_R}
+\frac{|\bv_{SR}^\dag\bT^{-1}\bw_n|^2}{\bv_{SR}^\dag\bT^{-1}\bv_{SR}}\right.
\\
\left.+\frac{|\bv_{S}^\dag\bT^{-1}\bw_m|^2}{\bv_{S}^\dag\bT^{-1}\bv_{S}}
\right\}\test \eta,
\label{eqn:EPGLRT1_CFAR_gamma}
\end{multline}
where $\bT=\sum_{k\in\Omega_S}\bt_k\bt_k^\dag$, showing that
it is invariant to data scaling operations. The same line of reasoning
can be also used for EP-GLRT-KM-2, EP-GLRT-KA, and A-GLRT to show their 
invariance with respect to $\gamma$. Thus, we omit the steps of the proof
except for the following key equalities that can be used for the proofs
\begin{align}
\frac{\bv_R^\dag\bS_S^{-1}\bz_1}{\bv_R^\dag\bS_S^{-1}\bv_R} &= 
{\gamma} \frac{\bv_R^\dag\bT^{-1}\bw_1}{\bv_R^\dag\bT^{-1}\bv_R},
\\
\frac{\bv_{SR}^\dag\bS_S^{-1}\bz_n}{\bv_{SR}^\dag\bS_S^{-1}\bv_{SR}} &=
{\gamma} \frac{\bv_{SR}^\dag\bT^{-1}\bw_n}{\bv_{SR}^\dag\bT^{-1}\bv_{SR}},
\\
\frac{\bv_S^\dag\bS_S^{-1}\bz_m}{\bv_S^\dag\bS_S^{-1}\bv_S} &=
{\gamma} \frac{\bv_S^\dag\bT^{-1}\bw_m}{\bv_S^\dag\bT^{-1}\bv_S},
\\
\frac{\bv_R^\dag\bS_{n,m}^{-1}\bz_1}{\bv_R^\dag\bS_{n,m}^{-1}\bv_R} &= 
{\gamma} \frac{\bv_R^\dag\bT_{n,m}^{-1}\bw_1}{\bv_R^\dag\bT_{n,m}^{-1}\bv_R},
\\
\frac{\bv_{SR}^\dag\bS_{n,m}^{-1}\bz_n}{\bv_{SR}^\dag\bS_{n,m}^{-1}\bv_{SR}} &=
{\gamma} \frac{\bv_{SR}^\dag\bT_{n,m}^{-1}\bw_n}{\bv_{SR}^\dag\bT_{n,m}^{-1}\bv_{SR}},
\\
\frac{\bv_S^\dag\bS_{n,m}^{-1}\bz_m}{\bv_S^\dag\bS_{n,m}^{-1}\bv_S} &=
{\gamma} \frac{\bv_S^\dag\bT_{n,m}^{-1}\bw_m}{\bv_S^\dag\bT_{n,m}^{-1}\bv_S},
\end{align}
where $\bT_{n,m}=\bT+\sum_{k\in\Omega_P\setminus \{1,n,m\}} \bw_k\bw_k^\dag$.

As for the C-GLRT, the proof is more involved due to the 
cyclic estimation procedures and
the CFAR behavior strictly depends on the initial values. Specifically,
let us write 
\be
\alpha_n^{(h-1)}={\gamma}u_n^{(h-1)}(\cD), \ \alpha_m^{(h-1)}={\gamma}u_m^{(h-1)}(\cD)
\label{eqn:condC-GLRT}
\ee
where $\cD=\{\bx_1,\ldots,\bx_{K_P},\bt_1,\ldots,\bt_{K_S}\}$,
then, we can write
\begin{align}
\bC_{n,m}^{(h-1)}&=\gamma^2\bT_{n,m}+\gamma^2 \left(\bW_{n,m}-\bV_{SRS} \bU_{n,m}^{(h-1)}\right)
\nonumber
\\
&\times \left(\bW_{n,m}-\bV_{SRS} \bU_{n,m}^{(h-1)}\right)^\dag
\nonumber
\\
&=\gamma^2\tilde{\bC}_{n,m}^{(h-1)},
\end{align}
where $\bW_{n,m}=[\bw_n,\bw_m]\in\C^{N\times 2}$,
the expression of $\tilde{\bC}_{n,m}^{(h-1)}$ can be easily obtained
from the above equation, and 
$\bU_{n,m}^{(h-1)}=\diag([u_n^{(h-1)}(\bW),u_m^{(h-1)}(\bW)]^T)\in\C^{2\times 2}$.
As a consequence, the update equation for $\alpha_1$ (step $1$) can be recast as
\be
\widehat{\alpha}_1^{(h)}=\gamma\frac{\bv_R^\dag(\tilde{\bC}_{n,m}^{(h-1)})^{-1}\bw_{1}}
{\bv_R^\dag(\tilde{\bC}_{n,m}^{(h-1)})^{-1}\bv_R}=\gamma u_1^{(h)}(\cD).
\ee
Focusing on step $2$ of the $h$th iteration, we observe that
\begin{align}
\bC_{1,m}^{(h-1)}&=\gamma^2\bT_{n,m}+\gamma^2 \left(\bW_{1,m}-\bV_{RS} \bU_{1,m}^{(h-1)}\right)
\nonumber
\\
&\times \left(\bW_{1,m}-\bV_{RS} \bU_{1,m}^{(h-1)}\right)^\dag
\nonumber
\\
&=\gamma^2\tilde{\bC}_{1,m}^{(h-1)},
\end{align}
where $\bW_{1,m}=[\bw_1,\bw_m]\in\C^{N\times 2}$ 
and $\bU_{1,m}^{(h-1)}=\diag([u_1^{(h)}(\cD),u_m^{(h-1)}(\cD)]^T)\in\C^{2\times 2}$,
and, hence, the estimate of $\alpha_n$ can be written as
\be
\widehat{\alpha}_n^{(h)}=\gamma\frac{\bv_{SR}^\dag(\tilde{\bC}_{1,m}^{(h-1)})^{-1}\bw_{n}}
{\bv_{SR}^\dag(\tilde{\bC}_{1,m}^{(h-1)})^{-1}\bv_{SR}}=\gamma u_n^{(h)}(\cD).
\ee
Finally, we apply the same line of reasoning to step $3$. Thus, 
we can write
\begin{align}
\bC_{1,n}^{(h)}&=\gamma^2\bT_{n,m}+\gamma^2 \left(\bW_{1,n}-\bV_{RSR} \bU_{1,n}^{(h)}\right)
\nonumber
\\
&\times \left(\bW_{1,n}-\bV_{RSR} \bU_{1,n}^{(h)}\right)^\dag
\nonumber
\\
&=\gamma^2\tilde{\bC}_{1,n}^{(h)},
\end{align}
where $\bW_{1,n}=[\bw_1,\bw_n]\in\C^{N\times 2}$ 
and $\bU_{1,n}^{(h)}=\diag([u_1^{(h)}(\cD),u_n^{(h)}(\cD)]^T)\in\C^{2\times 2}$.
Then, the estimate of $\alpha_m$ can be recast as
\be
\widehat{\alpha}_m^{(h)}=\gamma\frac{\bv_{S}^\dag(\tilde{\bC}_{1,n}^{(h)})^{-1}\bw_{m}}
{\bv_{S}^\dag(\tilde{\bC}_{1,n}^{(h)})^{-1}\bv_{S}}=\gamma u_m^{(h)}(\cD).
\ee
At the end of the iterative procedure, it is possible to show that the left-hand side
of \eqref{eqn:C-GLRT_final} is functionally independent of $\gamma$.
Thus, if \eqref{eqn:condC-GLRT} holds for $h=1$, then the C-GLRT is invariant
to the clutter power.

Now, let us focus on the structure $\bGamma$
and notice that proving the invariance with respect
to $\bGamma$ (at least to the best of authors' knowledge) is not an easy task due 
to the joint presence of different steering vectors
in the decision statistics. Nevertheless, we show below that
all the decision schemes are bounded CFAR meaning that
they are limited from above by a statistic that is invariant
with respect to $\bM$ (and, hence, $\bGamma$).
As a matter of fact, starting from the estimates and plug solutions, $\forall n,m\in\Omega_P$ 
the following inequality holds
\begin{multline}
\frac{|\bv_R^\dag\bS_S^{-1}\bz_1|^2}{\bv_R^\dag\bS_S^{-1}\bv_R}
+\frac{|\bv_{SR}^\dag\bS_S^{-1}\bz_n|^2}{\bv_{SR}^\dag\bS_S^{-1}\bv_{SR}}
+\frac{|\bv_S^\dag\bS_S^{-1}\bz_m|^2}{\bv_S^\dag\bS_S^{-1}\bv_S}
\\
\leq
\bz_1^\dag\bS_S^{-1}\bz_1+\sum_{i=n,m} \bz_i^\dag\bS_S^{-1}\bz_i
\end{multline}
and, hence, we have that
\begin{align}
&\dmax_{n,m\in\Omega_P \atop 
n>1,m>n} \left\{ \frac{|\bv_R^\dag\bS_S^{-1}\bz_1|^2}{\bv_R^\dag\bS_S^{-1}\bv_R}
+\frac{|\bv_{SR}^\dag\bS_S^{-1}\bz_n|^2}{\bv_{SR}^\dag\bS_S^{-1}\bv_{SR}}
+\frac{|\bv_S^\dag\bS_S^{-1}\bz_m|^2}{\bv_S^\dag\bS_S^{-1}\bv_S}
\right\}
\nonumber
\\
&\leq
\dmax_{n,m\in\Omega_P \atop 
n>1,m>n} \left\{ \bz_1^\dag\bS_S^{-1}\bz_1+\sum_{i=n,m} \bz_i^\dag\bS_S^{-1}\bz_i\right\}.
\label{eqn:boundedCFAR_EP-GLRT-KM-1}
\end{align}
Now, notice that the statistical characterization of $\bz_l^\dag\bS_S^{-1}\bz_l$, $l\in\{1,n,m\}$, under $H_0$,
does not depend on $\bM$. In fact it can be written as $\bz_{l,0}^\dag\bS_{S,0}^{-1}\bz_{l,0}$,
where $\bz_{l,0}=\bM^{-1/2}\bz_{l}\sim\cC\cN_N(\bzero,\bI)$ and 
$\bS_{S,0}=\bM^{-1/2}\bS_{S}\bM^{-1/2}\sim\cC\cW_N(K_S,\bI)$ are random quantities independent
of $\bM$. It follows that the right-hand side of \eqref{eqn:boundedCFAR_EP-GLRT-KM-1} under $H_0$
is invariant with respect to $\bM$ (and, hence, to $\bGamma$). This property can be similarly proved
also for the EP-GLRT-KM-2 when $\bS_S$ is replaced by $\bS_{n,m}$.

As for the EP-GLRT-KA \eqref{eqn:EP-GLRT-KA}, the C-GLRT \eqref{eqn:C-GLRT_final}, and
the A-GLRT \eqref{eqn:A-GLRT_final}, we exploit the inequality \cite{lutkepohl1997handbook}
\be
\det(\bA+\bB)\geq \det(\bA),
\ee
where $\bA\in\C^{N\times N}$ is positive definite and $\bB\in\C^{N\times N}$
is positive semidefinite, to obtain
\be
\left.\begin{matrix}
\dmax_{n,m\in\Omega_P \atop 
n>1,m>n}\frac{\det(\bS_P+\bS_S)}
{\det[\bS_{n,m} + \bY_{1,n,m}]} 
\\
\vspace{-1mm}
\\
\dmax_{n,m\in\Omega_P \atop 
n>1,m>n}\frac{\det(\bS_P+\bS_S)}
{\det\left(\bS_{n,m} + \bar{\bY}_{1,n,m}\right)}
\\
\vspace{-1mm}
\\
\dmax_{n,m\in\Omega_P \atop 
n>1,m>n}\frac{\det(\bS_P+\bS_S)}
{\det\left[
\bS_{n,m} + \bQ_{n,m}
\right]}
\end{matrix}
\right\}
\leq
\dmax_{n,m\in\Omega_P \atop 
n>1,m>n}\frac{\det(\bS_P+\bS_S)}
{\det[\bS_{n,m}]}.
\ee
Now, the right-hand side of the previous equation can be recast as
\begin{multline}
\dmax_{n,m\in\Omega_P \atop 
n>1,m>n}\frac{\det(\bS_{P,0}+\bS_{S,0})\det(\bM)}
{\det[\bS_{n,m,0}]\det(\bM)}
\\
=\dmax_{n,m\in\Omega_P \atop 
n>1,m>n}\frac{\det(\bS_{P,0}+\bS_{S,0})}
{\det[\bS_{n,m,0}]},
\end{multline}
where $\bS_{P,0}=\sum_{k=1}^{K_P}\bM^{-1/2}\bz_k\bz_k^\dag \bM^{-1/2}$ and
$\bS_{n,m,0}=\bM^{-1/2}\bS_{n,m}\bM^{-1/2}\sim \cC\cW_N(K_S+K_P-3,\bI)$.
Again, neither the statistical characterization of $\bS_{P,0}$, $\bS_{n,m,0}$,
and $\bS_{S,0}$ or their correlation depend on $\bM$ and, hence,
the right-hand side of the last equation is invariant with respect to $\bM$.


\section{RCS of Low-RCS Aircrafts and RIS Design Issues}
\label{Sec:Low-RCS}
The purpose of this section is twofold. First, we start with the description
of the main design principles used to reduce the RCS of an aircraft also
providing examples of practical value. Then, 
we proceed by indicating some guidelines to develop a RIS tailored for the specific application.

\subsection{Low-RCS Aircrafts: Examples}
The RCS of an aircraft is not a single number but a complex 
signature dictated by four fundamental pillars. Shape is the paramount factor, where 
specific geometries are designed to deflect radar waves away from the source. 
Materials play a crucial supporting role, with radar-absorbing materials (RAM) 
converting incident radar energy into heat to reduce reflection. 
Intrinsic reflectivity involves managing the inherent properties of surfaces and, 
critically, concealing cavity reflectors like engine inlets and cockpits, 
which are major sources of radar returns. Finally, aspect angle acknowledges that 
an aircraft's RCS is highly dynamic, leading designers to strategically 
optimize stealth for a specific ``threat sector,'' most commonly 
the frontal aspect, accepting higher signatures from less critical angles.

The practical application of RCS reduction principles can be best understood 
by examining iconic stealth aircraft, each representing a different evolutionary stage and design philosophy.
The evolution of stealth aircraft demonstrates a progression in design philosophy and technology. 
The F-117 Nighthawk pioneered operational stealth by using 
faceted geometry to deflect radar waves in specific directions, 
a calculable but aerodynamically inefficient method effective from frontal angles.
The B-2 Spirit represented a major advance with its flying-wing shape 
and continuous curves, creating an all-aspect stealth design 
that scatters radar energy weakly and provides a consistently low 
signature from any direction.
Finally, the F-22 Raptor and F-35 Lightning II synthesize stealth with high performance. 
They use a blend of planform alignment and curved surfaces, heavily optimized 
for a low frontal RCS for air dominance. Key features include internal 
weapon bays and concealed engine inlets, with the F-35 incorporating 
advanced materials for better sustainability.
The F-35 is designed to significantly reduce radar returns, particularly 
in the X-band frequency range, where most fire-control and targeting radars operate
\cite{guan2025electromagnetic,herda2020radar,gurel2003validation,chung2016radar,alves2007simulations}.
These measures are most effective in the frontal hemisphere, where the 
aircraft presents minimal surface discontinuities and optimized planform 
alignment, resulting in an RCS as low as $-40$ dBsm under ideal nose-on conditions \cite{kopp2010evolving}.
In Table \ref{tab:rcs_cases}, we show the estimated RCS values for
dim airborne targets that fly at 
a low altitude of $500$ meters and is viewed from a range of $40$ km; the 
radar sees the aircraft at an elevation angle 
of approximately $0.72^\circ$, revealing only 
its nose.
This represents the stealth-optimal geometry, yielding 
an estimated RCS below $-30$ dBsm.

However, as the distance to the radar decreases, the elevation angle increases, 
exposing less stealth-optimized features on the lower fuselage, including the weapons bay doors, 
landing gear covers, and the trailing edges near the engine nozzle. 
At $10$ km, the elevation angle increases to approximately $2.86^\circ$, and the 
estimated RCS rises to about $-20$ dBsm. When the range decreases further to $5$ km, 
with an elevation angle of $5.71^\circ$, the radar has a clearer line 
of sight to the aircraft underbody, resulting in a further increase in RCS to the range $[-15,-20]$ dBsm. 
These values, while still significantly lower than those of conventional 
fourth-generation fighters (which typically exhibit RCS values between $1$ and $5$ m$^2$ 
in frontal aspects), demonstrate how even small shifts in viewing geometry can 
affect the radar observability of a stealth platform \cite{knott2004radar}. 

\begin{table*}[ht]
\centering
\caption{Monostatic and bistatic RCS estimates for low-RCS aircrafts in S-band (2-4 GHz)}
\label{tab:rcs_cases}
\resizebox{1\textwidth}{!}{
\begin{tabular}{ c c c c p{6cm} }
\hline
\textbf{Case} & \textbf{Geometry} & \textbf{Estimated RCS} & \textbf{Notes} \\
\hline
RTR (radar-target-radar) & Monostatic, nose-on ($\theta_i \approx 0^\circ$) & \SIrange{-30}{-20}{dBsm} & Best-case stealth performance. \\
STS (surface-target-surface) & Monostatic, $\theta_{So} \approx 30^\circ$ off-nose & \SIrange{-10}{0}{dBsm} & Monostatic from RIS \\
RTS (radar-target-surface) & Bistatic ($\theta_{So} \approx 30^\circ$), Rx at nose ($\theta_i \approx 0^\circ$) & \SIrange{-10}{0}{dBsm} & Bistatic path \\
\hline
\end{tabular}
}

\smallskip
\footnotesize
\textbf{Additional notes:}
\begin{itemize}
\itemsep-0.3em 
\item All values assume clean aircraft configuration (no weapons/doors open)
\item Bistatic advantage is less pronounced in S-band than X-band but still measurable
\item Worst-case side aspect (90$^\circ$) could reach \SIrange{-5}{+10}{dBsm} in S-band
\end{itemize}
\end{table*}

\subsection{Case Study, Energy Considerations, and RIS Design}
\label{subsec:operatingScenario}
\begin{figure*}
\begin{center}
\includegraphics[scale=0.25]{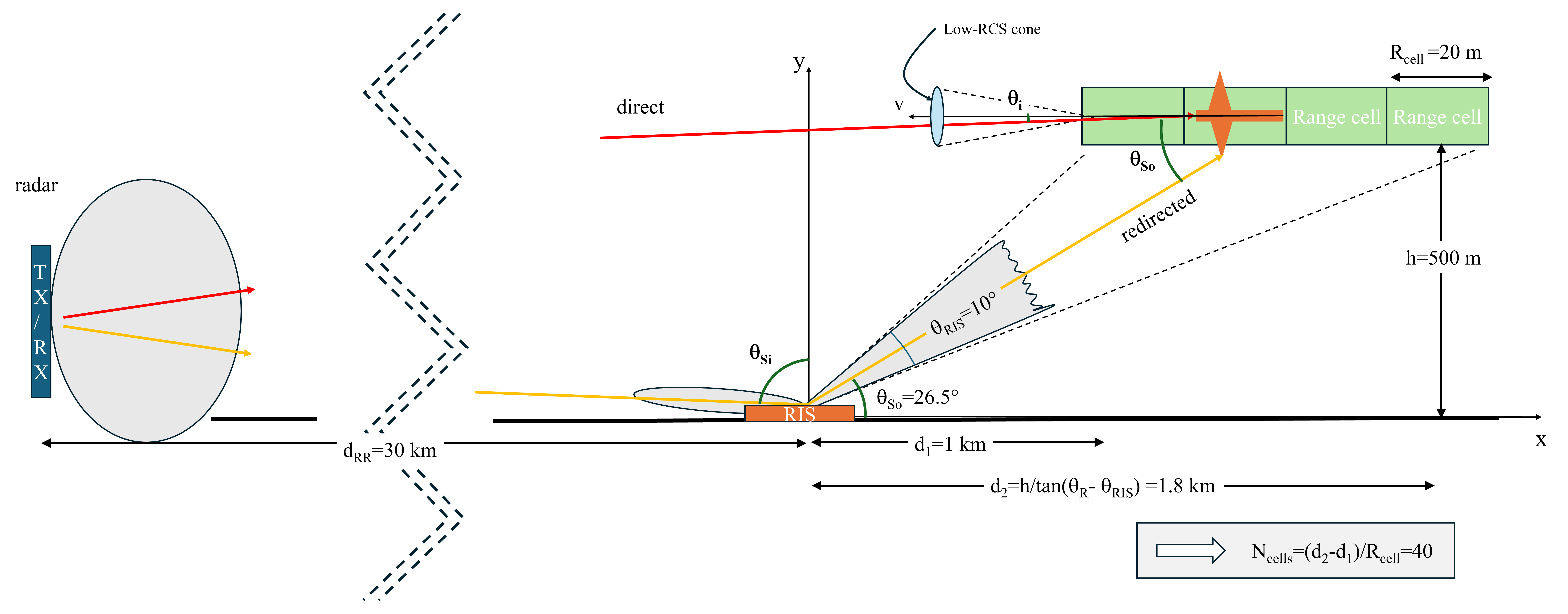}
\caption{Geometry of the investigated operating scenario.}
\label{fig_operatingScenarioXZ}
\end{center}
\end{figure*}

\begin{figure}
\begin{center}
\includegraphics[width=8cm]{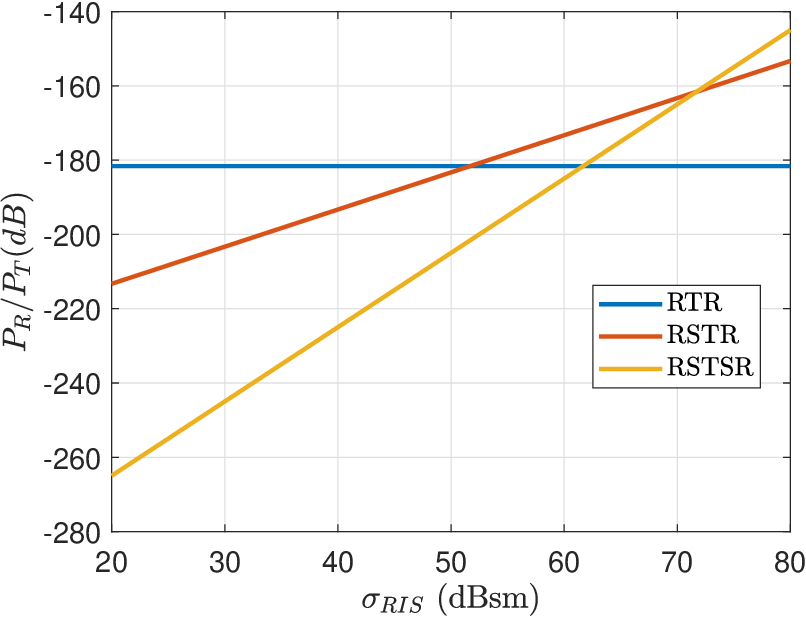}
\caption{Received power as a function of the RIS RCS for the three analyzed contributions.}
\label{fig_PR}
\end{center}
\end{figure}

The investigated scenario is depicted in Figure \ref{fig_operatingScenarioXZ}.
Specifically, given the reference system in the figure, 
the positions of the radar system, the RIS, and the target are the following
\begin{itemize}
    \item radar location: \( (-30000, \, 200) \) m;
    \item RIS location: \( (0,\, 0) \) m;
    \item target location: \( (1000,\, 500) \) m.
\end{itemize}
In this context, the relevant distances are
\be
\begin{cases}
d_{RT} = \sqrt{31^2 + 0.3^2}\approx 31\,\text{km}, 
\\
\vspace{-3mm}
\\
d_{RS} = \sqrt{30^2 + 0.2^2}\approx 30 \,\text{km},
\\
\vspace{-3mm}
\\
d_{ST} = \sqrt{1^2 + 0.5^2}\approx 1.2 \,\text{km},
\end{cases}
\ee
and, assuming a range resolution of $20$ m, it is not difficult 
to verify that the above values satisfy conditions \eqref{eqn:cond_tau}. 
Based on the above parameters and distances, we can compute the incident and reflection angles 
relative to the RIS, which is placed at the origin of the reference system. 
The angle \(\theta_{\text{Si}}\) is measured between the incoming 
wave (from radar to RIS) and the RIS normal (vertical axis) and
is given by
\be
\theta_{\text{Si}} = \tan^{-1}\left( \frac{30000}{200} \right) \approx 89.62^\circ.
\ee
The angle \(\theta_{\text{So}}\) between the RIS-to-target line and
the $x$-axis is
\be
\theta_{\text{So}} = 90^\circ-\tan^{-1}\left( \frac{1000}{500} \right) \approx 26.5^\circ.
\ee
The simulation results shown in the next section assume that
the carrier frequency of the radar is $f_c=3$ GHz which corresponds 
to a wavelength of $10$ cm. Moreover, the transmitted
power is $P_T=10$ kW and the transmit gain is $G_T = 37$ dBi,
which means an effective area of the transmitting/receiving
antenna being \cite{balanis2016antenna}
\be
A_{eff} = \frac{\lambda^2 G_T}{4\pi}.
\ee
In order to compute the received energy according to the different paths, we need to know the values of the 
target monostatic and bistatic RCSs. Specifically, the target RCS values considered are
\begin{itemize}
\item LOS path: $\sigma_{RTR} = 10^{-2}\,\text{m}^2 $ (i.e., -20 dBsm);
\item RIS-RIS path: $ \sigma_{STS} = 1\,\text{m}^2 $ (i.e., 0 dBsm);
\item RIS-assisted LOS: $\sigma_{STR}= 1\,\text{m}^2 $ (i.e., 0 dBsm).
\end{itemize}
Thus, the received power for the different paths as a function of the RIS RCS is
\begin{itemize}
\item LOS path:
\be
P_{R}^{RTR}(\sigma_{RIS}) = \frac{P_T \, G_T \, \sigma_{RTR} \, A_{eff}}{(4\pi)^2 \, d_{RT}^4};
\label{LOS_equation}
\ee
\item RIS-assisted LOS Path (single bounce):
\begin{multline}
P_{R}^{RSTR}(\sigma_{RIS}) 
= \frac{P_T \, G_T}{4\pi d_{RS}^2}  
\\
\times \sigma_{RIS}  
  \frac{1}{4\pi d_{ST}^2}  \sigma_{STR}  
  \frac{A_{eff}}{4\pi d_{RT}^2};
\label{single_bounce_equation}
\end{multline}
\item RIS-RIS Path (double bounce):
\begin{multline}
P_{R}^{RSTSR} (\sigma_{RIS})
= \frac{P_T \, G_T}{(4\pi d_{RS}^2)^2}  
\\
\times \sigma_{RIS}^2  
  \frac{1}{(4\pi d_{ST}^2)^2}  \sigma_{STS} A_{eff}.
\label{double_bounce_equation}
\end{multline}
\end{itemize}
In Figure \ref{fig_PR}, we plot the ratio of the received power over transmitted power
as a function of the RIS RCS $\sigma_{RIS}$, which sweeps from $10$ 
dBsm to $80$ dBsm,
for the three paths by using the aforementioned parameter values. 
The curve obtained from \eqref{LOS_equation} is constant 
as it does not depend on $\sigma_{RIS}$.
The curves obtained from \eqref{single_bounce_equation} 
and \eqref{double_bounce_equation} indicate that 
in order to have a relevant contribution 
from secondary paths, the RIS should be designed with an RCS larger 
than $55$ dBsm.

Now, we have all the elements to proceed with the design of the RIS. As a matter of fact, 
the above analysis draws some guidelines for the design of the RIS and, in particular, allows us 
to establish the minimum RIS aperture $L_{\text{RIS}}$ such that the RIS-assisted path dominates 
the LOS path. To this end, we model the RIS as 
a square aperture of size $L_{\text{RIS}} \times L_{\text{RIS}}$. 
It is important to underline that a key advantage 
of a RIS (or, more generally, 
a metasurface) is its ability to control the aperture illumination profile, 
or tapering, either passively or actively. By manipulating 
the phase and amplitude of each unit cell, the reflected wavefront can be shaped to realize 
a desired tapering across the aperture, which directly impacts the effective RCS, 
beamwidth, and sidelobe levels.  

Once an optimal phase or tapering profile is determined (e.g., to steer reflected 
energy toward a target at a desired 
bistatic angle) the RIS configuration can be synthesized 
using a local periodicity approximation, 
a standard method in reflectarray antenna design \cite{pozar1997design, borgese2017iterative}. 
This approach allows the aperture illumination (tapering) to be tailored precisely, either 
statically or dynamically, providing agile control over beam shaping and 
effective RCS for stealthy or multi-angle observable targets.

The RIS can be also designed as reconfigurable surface. Active control is typically achieved 
using tunable elements such as varactors \cite{sievenpiper2003two, costa2008active}, 
PIN diodes \cite{yu2022reconfigurable}, or liquid crystals \cite{aghabeyki2023dual}. 
Varactor diodes are especially attractive for radar RIS applications due 
to their continuous phase tuning, low power requirements, and high-speed response, 
enabling rapid adaptation to dynamic scenes. However, they require more complex analog biasing networks. 
PIN diodes simplify the circuitry as binary switches but offer coarse phase resolution. 
Liquid crystal-based RIS are promising for large-area and mm-wave applications, though 
their slower switching speed limits real-time radar use.

At the design stage, three illumination functions are considered along the $x$-axis: 
uniform illumination, sinc tapering, and linear frequency modulated (LFM) phase-only tapering 
(with uniform distribution along $y$). While uniform illumination maximizes 
the RIS RCS, it does not provide control over the beamwidth, which becomes increasingly 
narrow as the RIS size grows. This narrow beam is not compatible with the simultaneous illumination of
multiple range cells. To address this, sinc and LFM tapering are investigated along 
the $x$-axis to achieve 
a controlled beamwidth over a predefined angular sector, allowing the designer 
to distribute energy across multiple targets or observation directions
while maintaining an acceptable loss in terms of RIS RCS.

\subsubsection{$2$-dimensional Uniform Illumination}
Considering a square surface with side length $L_{\text{RIS}}$, 
it is possible to show that the boresight RCS is:
\be
\sigma_{\text{uniform}} = \frac{4\pi L_{\text{RIS}}^4}{\lambda^2}.
\ee
This corresponds to the maximum RCS achievable for a perfectly phased square aperture. This implies that the minimum RIS size that allows to get a relevant contribution from secondary paths is
\be
L_{\text{RIS}} = \left( \frac{\sigma_{RIS} \, \lambda^2}{4\pi} \right)^{1/4}
\ee
Normalizing $L_{\text{RIS}}$ with respect to wavelength, and assuming a periodicity of half-wavelength for the RIS elements, we get the number of elements per line, $N_{RIS}$:
\be
N_{\text{RIS}} = \left\lceil 2  \frac{L_{\text{RIS}}}{\lambda} \right\rceil.
\ee

The Half Power Beamwidth (HPBW) of a uniform linear RIS array can then be approximated as \cite{balanis2016antenna}:
\be
\text{HPBW} \approx \frac{2 \times 50.8^\circ}{N_{\text{RIS}}},
\ee
which provides an estimate of the spatial resolution achievable with the RIS. By employing the design parameters, a HPBW angle of $1.5^{\circ}$ is obtained. However, since a larger angle is needed to 
illuminate multiple radar cells (e.g., $10^{\circ}$) a sinc tapering of the surface can be employed.

\subsubsection{Sinc Tapering along the x-axis and Uniform Illumination along y-axis}
\begin{figure}
\begin{center}
\includegraphics[width=0.85\linewidth]{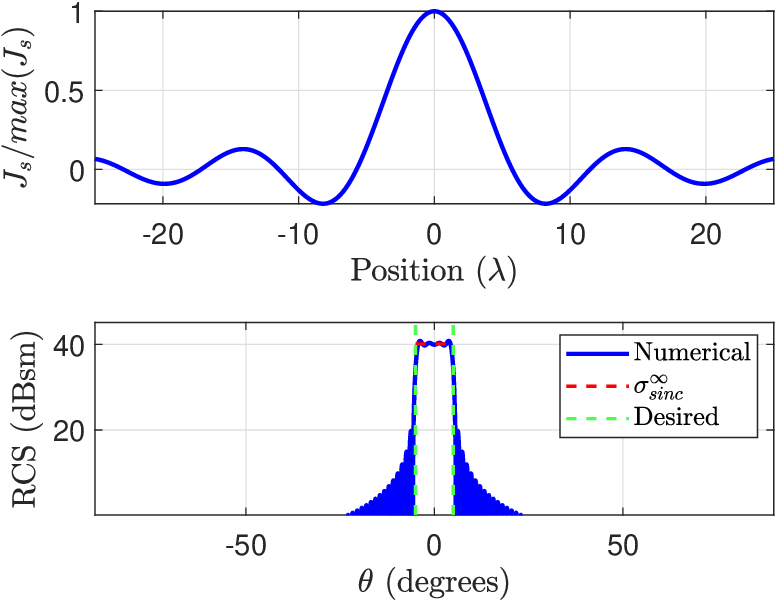}
\caption{Sinc tapering of the aperture and RCS pattern obtained for an aperture of $50\lambda$.}
\label{fig_sinc_tapering_pattern}
\end{center}
\end{figure}

\begin{figure}
\begin{center}
\includegraphics[width=0.85\linewidth]{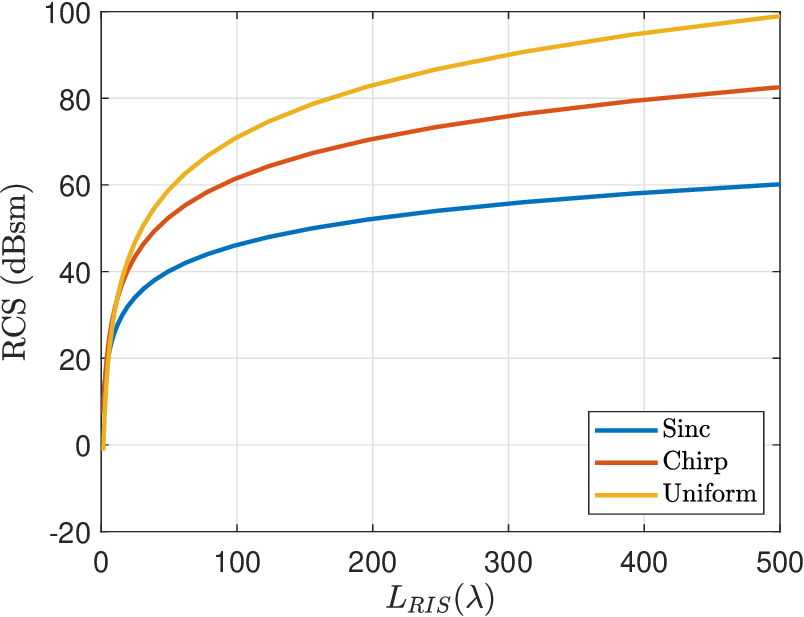}
\caption{Predicted RCS of a RIS with uniform and Sinc and LFM tapering for $\phi_0=10 ^{\circ}$.}
\label{fig_comparison_RCS}
\end{center}
\end{figure}

The surface current distribution imposed by the RIS can be written as
\be
J_s(x,y) = 2 \frac{E_0}{\zeta_0} \text{sinc}\!\left(\frac{x}{b}\right) \text{rect}\!\left(\frac{x}{L_{\text{RIS}}}\right)
\text{rect}\!\left(\frac{y}{L_{\text{RIS}}}\right),
\ee
where $E_0$ is the impinging electric field, $\zeta_0=\sqrt{\mu_0/\varepsilon_0}$ 
is the free space impedance, where $\varepsilon_0$ and $\mu_0$ represent 
the vacuum dielectric permettivity and permeability, respectively. 
The parameter $b$ represents the tapering and controls the mainlobe width 
and sidelobes' level; its definition is $b=\lambda/\phi_0$,
where $\phi_0$ is the desired beamwidth.
An example is shown in Figure \ref{fig_sinc_tapering_pattern} where a sinc tapering is 
applied to an aperture of 50 wavelenghts and a rectangular pattern with the desired beamwidth 
($10^{\circ}$) is obtained. The beamwidth control is paid in terms of 
maximum RCS of the aperture as corroborated by Figure \ref{fig_comparison_RCS}. 

At boresight, the RCS of the sinc-tapered aperture becomes:
\begin{equation}
\sigma_{\text{sinc}} = \frac{16 b^2 L_{\text{RIS}}^2}{\pi \lambda^2} \,
\text{Si}^2\!\left(\frac{\pi L_{\text{RIS}}}{2b}\right),
\end{equation}
with $\text{Si}(z) = \int_0^z \frac{\sin t}{t} dt$.
For a very large aperture ($L_{\text{RIS}} \to \infty$):
\begin{equation}
\sigma_{\text{sinc}}^{\infty} = \frac{4\pi b^2 L_{\text{RIS}}^2}{\lambda^2}.
\end{equation}
%
%
\begin{figure}
\begin{center}
\includegraphics[width=0.85\linewidth]{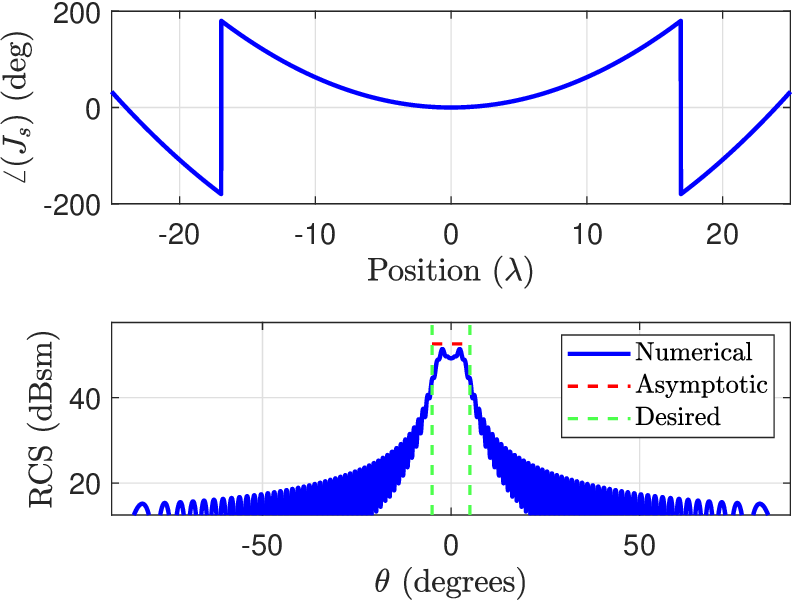}
\caption{LFM phase-only tapering of the aperture and RCS pattern obtained for an aperture of $50\lambda$.}
\label{fig_chirp_tapering_pattern}
\end{center}
\end{figure}
%
\subsubsection{LFM Phase-only Tapering along the x-axis and Uniform Illumination along y-axis}
In case of LFM tapering \cite{farina2025efficient}, the RCS reduction is less 
critical than Sinc tapering and an acceptable beam control is allowed.
An LFM-tapered RIS can be modeled by imposing a quadratic phase distribution 
along the $x$-axis, while maintaining uniform illumination along $y$. 
The resulting surface current density can be written as
\be
J_s(x,y) = 2 \frac{E_0}{\zeta_0}
\, e^{j\pi K_x x^2}\text{rect}\!\left(\frac{x}{L_{\text{RIS}}}\right)
\, \text{rect}\!\left(\frac{y}{L_{\text{RIS}}}\right),
\ee
where $K_x$ is the chirp rate that controls the angular spreading of 
the scattered field. In particular, for a rectangular beam of 
angular width $\phi_0$ in the $x$--$z$ plane, the chirp rate can be approximated as
\be
K_x \simeq \frac{\phi_0}{\lambda L_{\text{RIS}}},
\ee
which directly links the phase curvature to the desired beamwidth. 
The quadratic phase prevents coherent addition 
over the full aperture and limits the effective contributing 
region along the $x$-axis to a stationary-phase length, thereby redistributing 
the scattered energy over a broader angular sector, as illustrated 
in Figure \ref{fig_chirp_tapering_pattern}.
At boresight, the RCS of the LFM-tapered aperture can be asymptotically estimated 
using the stationary-phase method as
\be
\sigma^{\infty}_{\text{chirp}} \approx \frac{4\pi}{\lambda^2}
\left( L_{\text{RIS}} \sqrt{\frac{2}{K_x}} \right)^2
=\frac{8\pi}{\lambda^2} \frac{L_{\text{RIS}}^2}{K_x}.
\ee
Thus, for a very large aperture, the peak RCS scales linearly with the aperture dimension along 
the LFM-tapered direction,
highlighting the fundamental tradeoff between angular beamwidth 
control and maximum achievable RCS in LFM-tapered RIS apertures.

\medskip

The comparison of the RIS RCSs corresponding to the aforementioned tapering approaches 
is reported in Figure \ref{fig_comparison_RCS}.
The following concluding remarks can be drawn
\begin{itemize}
\item the uniform illumination maximizes the RCS, which scales 
as $L_{\text{RIS}}^4$ but it does not provide any beamwidth control;
\item Sinc tapering effectively reduces the coherent aperture along 
the $x$-direction, resulting in a lower peak RCS while enabling improved beam 
shaping and sidelobe suppression; the tapering parameter $b$ controls 
the mainlobe width and the effective RCS: smaller values of $b$ lead to wider beams at the expense of reduced peak RCS;
\item LFM phase-only tapering represents an attractive compromise: although sidelobe 
control is suboptimal compared to amplitude tapering, it enables high RCS levels 
(exceeding $60$ dBsm for a side length of $100\,\lambda$) while providing flexible 
angular beam shaping through phase-only modulation.
\end{itemize}

\section{Performance Assessment}
\label{Sec:PerformanceAnalysis}
In this section, we assess the performance of the proposed detection schemes by
generating target echoes according to the scenario described in Subsection \ref{subsec:operatingScenario}.
The performance metric is the $P_d$ (for a preassigned
value of the $P_{fa}$) as a function of the SINR for 
the monostatic path (RTR) defined as
\be
\mbox{SINR}_{RTR}=|\alpha_1|^2 \bv_R\bM^{-1}\bv_R.
\label{eqn:SINR}
\ee
The above metric is estimated by resorting to 
standard Monte Carlo (MC) counting techniques. Specifically, we compute the
detection thresholds and the $P_d$ values over $100/P_{fa}$ 
and $10^4$ independent trials, respectively. As for the radar system, we 
consider a uniform linear array formed by $N$ vertically aligned sensors, 
whose inter-element spacing is equal to $\lambda/2$ with $\lambda$ 
the operating wavelength and assume that only spatial processing 
is performed (for simplicity). The corresponding steering vector can be expressed as 
$\bv(\theta)= \left[1, e^{j\pi\sin\theta},\ldots,
e^{j\pi(N-1)\sin\theta}\right]^T\in\C^{N\times 1}$,
where $\theta$ is the elevation angle. Based on 
Figure \ref{fig_operatingScenarioXZ}, we set $\theta=0.5^\circ$ for $\bv_R$  
and $\theta=-0.4^\circ$ for $\bv_S$; moreover, since only spatial 
channels are considered in the simulation setup, 
it follows that $\bv_{S,1}=\bv_S$ and $\bv_{R,1}=\bv_R$.

According to the design guidelines defined in the previous section, we consider
LFM phase-only beamforming and a RIS whose size allows the bistatic paths 
to yield echoes with powers that are about $20$ dB stronger
than those from the monostatic path. Thus, once $\alpha_1$ is set according to \eqref{eqn:SINR},
we consider $\alpha_n=\alpha_m=10\alpha_1$. Moreover, let us recall that the range 
resolution is $20$ m and assume that the region under 
surveillance is formed by $20$ out of $40$ available range bins 
as indicated in Figure \ref{fig_operatingScenarioXZ}.
Thus, exploiting the distances mentioned in Subsection \ref{subsec:operatingScenario}
and assuming, for simplicity, that the monostatic echo 
occupies the first range bin,
it is possible to show that the range bins containing the returns
from the bistatic paths are indexed by $n=3$ for the path RTSR (or RSTR) and $m=6$ for the path RSTSR.
With these values in mind, we set $K_P=6$.\footnote{It is clear that this system parameter depends
on the geometry of the operating scenarios and a suitable set of guard cells can be added to account
for scenario uncertainty.}

As for the interference component, it is modeled as a complex Gaussian
random vector with covariance matrix $\bM=\sigma^2_n\bI+\bM_c$, where
$\sigma^2_n=1$ is the noise power and $\bM_c$ is the clutter
covariance matrix whose $(n,m)$th entry 
is $\sigma^2_c \rho^{|n-m|}$ with $\rho=0.9$ the one-lag
correlation coefficient and $\sigma^2_c>0$
the clutter power set according to 
the clutter-to-noise ratio (CNR) defined as 
$\mbox{CNR}=10\log_{10}(\sigma^2_c/\sigma^2_n)=25$ dB.
Finally, in the following numerical examples, we assume $N=16$ and $P_{fa}=10^{-4}$.

\subsection{Convergence of the C-GLRT}
\begin{figure}
\begin{center}
\centering  
\subfloat[$K_S=24$]{\label{fig:Convergence24}
\includegraphics[width=0.83\linewidth]{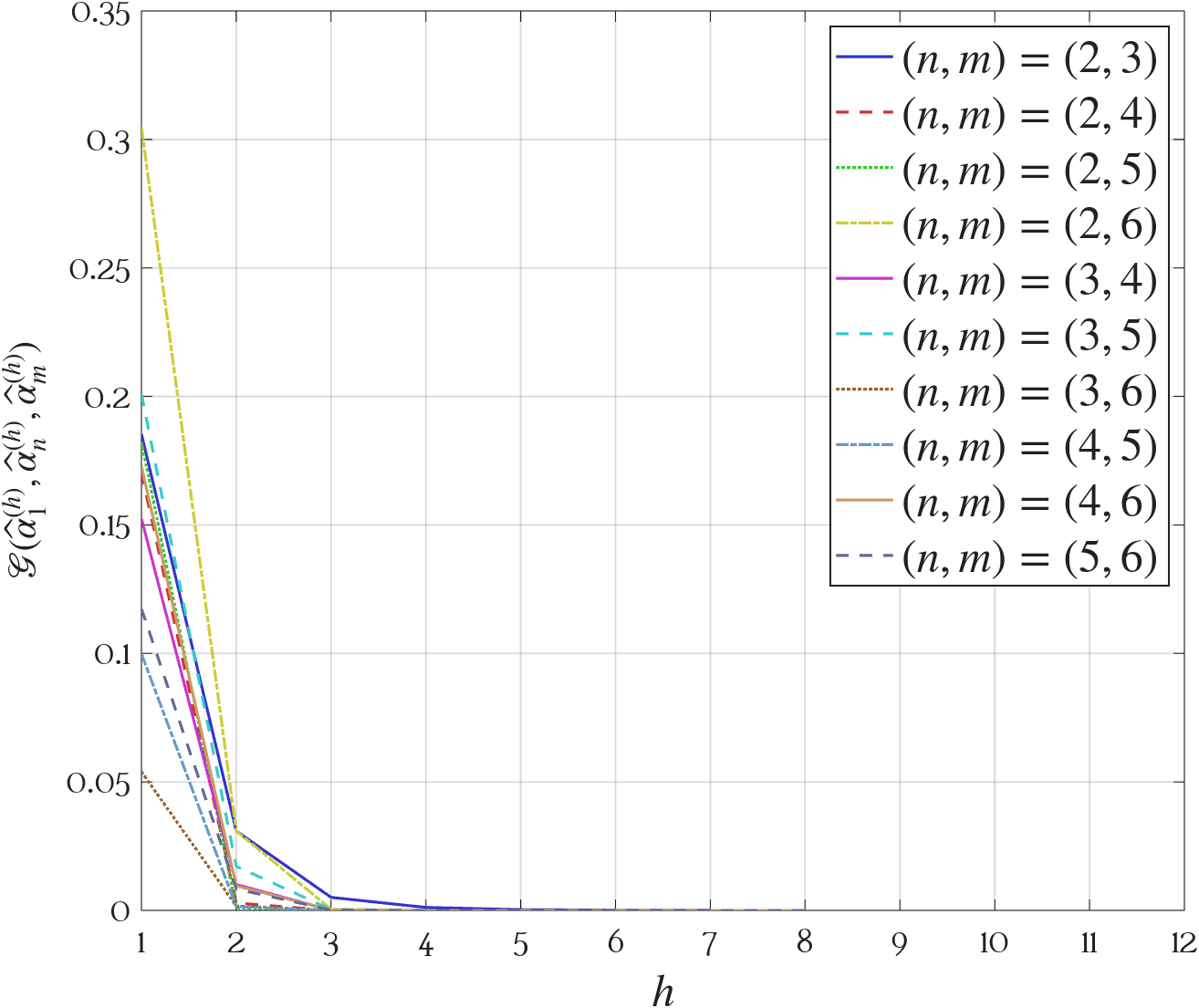}}
\\
\subfloat[$K_S=32$]{\label{fig:Convergence32}
\includegraphics[width=0.83\linewidth]{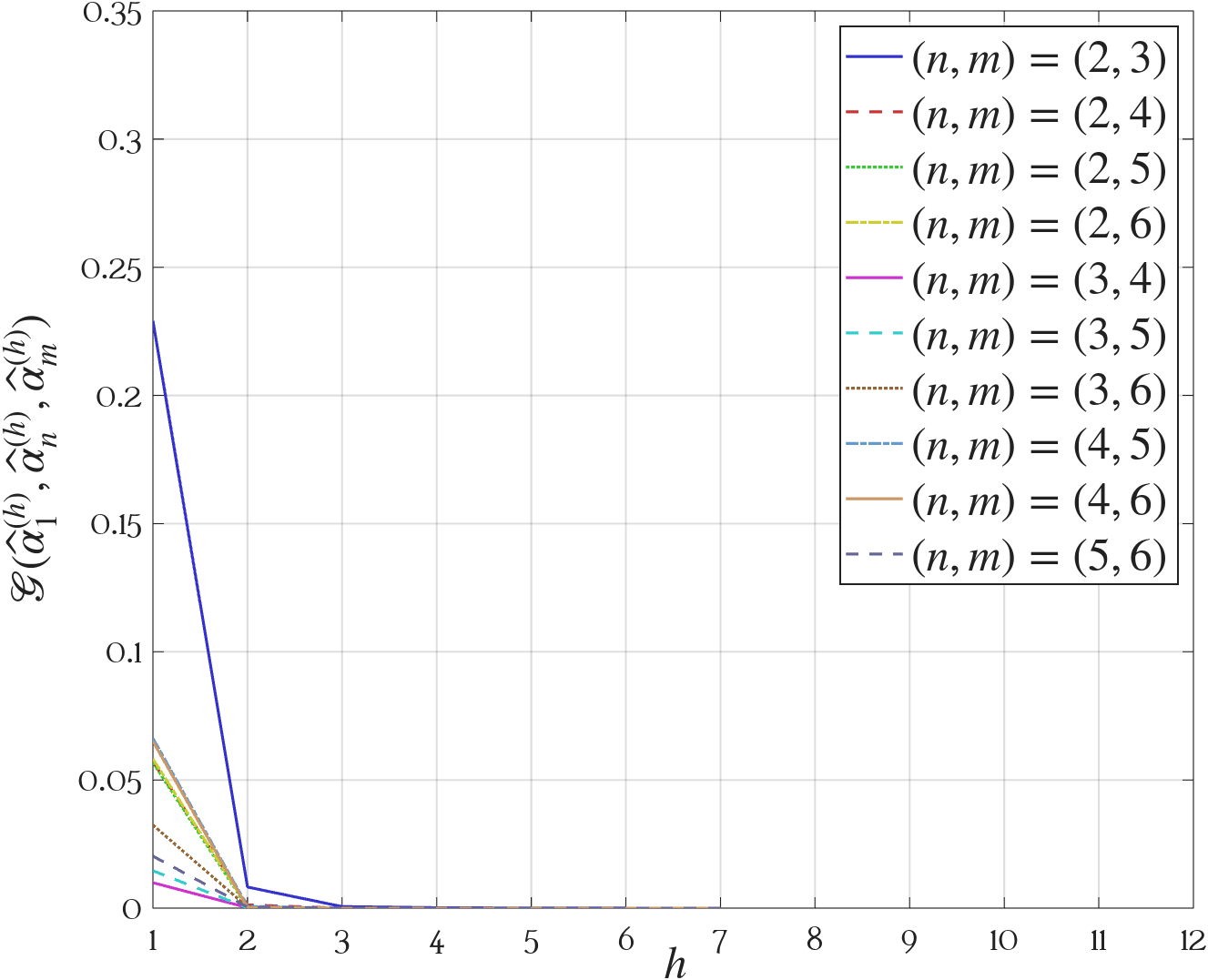}}
\caption{Convergence analysis of the C-GLRT assuming $N=16$ 
and different pairs $(n,m)$; (a) $K_S=24$, (b) $K_S=32$.}
\label{fig:Convergence}
\end{center}
\end{figure}

This subsection verifies the convergence rate of the 
C-GLRT algorithm. To this end, in Figure \ref{fig:Convergence} we 
plot the left-hand side of \eqref{eqn:stoppingCriterion},
namely the relative variation of the function $g(\cdot)$ defined in 
Subsection \ref{subsec:C-GLRT}, averaged over $1000$ MC trials
versus the iteration index $h$ and under $H_1$ when the target geometry generates
different $(n,m)$ pairs. In the numerical example, 
the training sample size is $K_S=24,32$. 
The figure shows that all the curves drop to very low values after some iteration. 
As a matter of fact, in \eqref{eqn:stoppingCriterion}, 
if we set $\epsilon=10^{-5}$, 
the inequality is generally satisfied for $h \le 8$. For this reason, in what follows, 
we use the above value for $\epsilon$ and we set $h_{\max}=20$.

\subsection{CFAR Property}
\label{Sec:CFARproperty}
\begin{figure}[t]
\centering  
\subfloat[$P_{fa}$ versus CNR]{\label{fig:CFAR_CNR}
\includegraphics[width=0.87\linewidth]{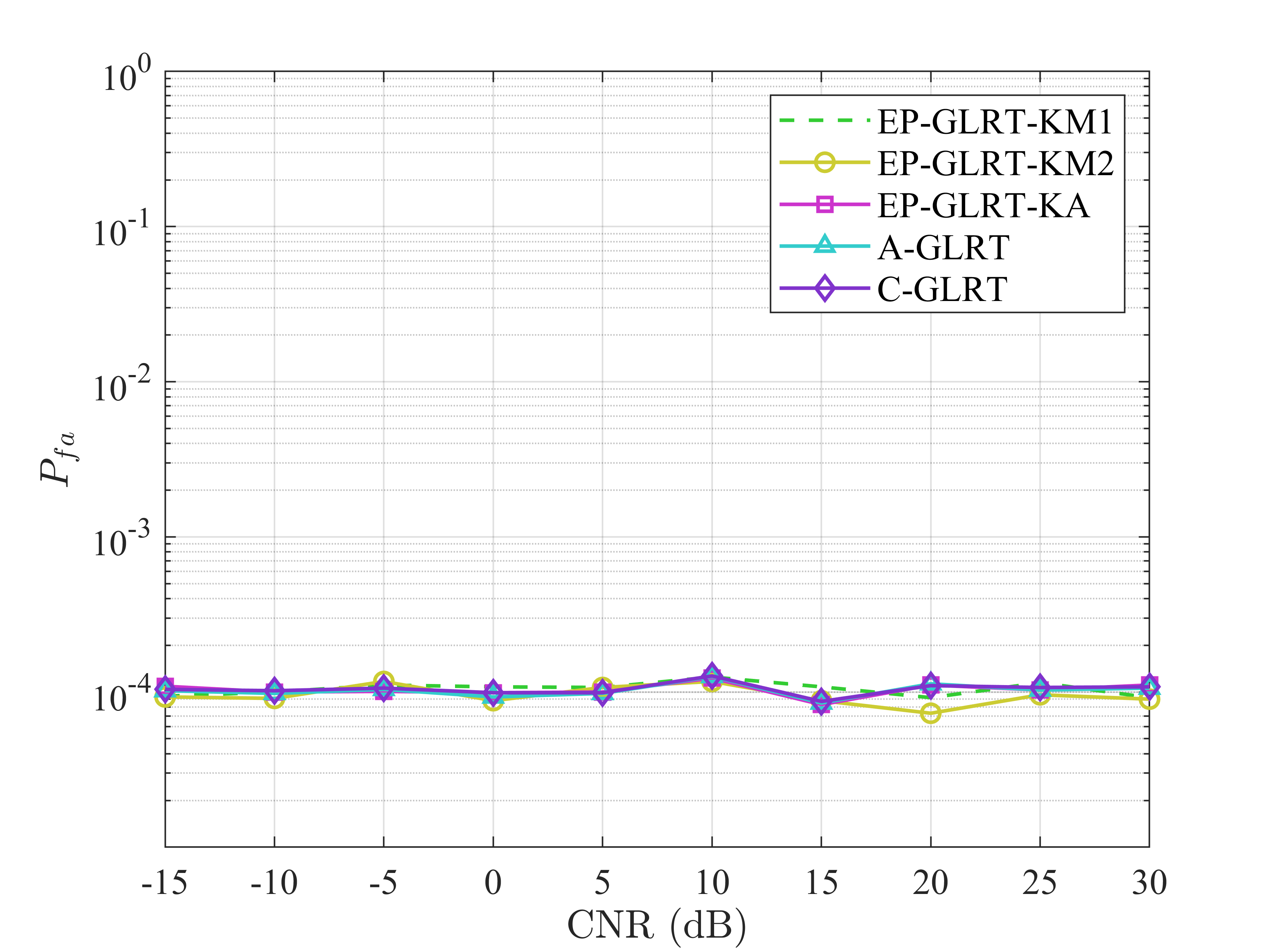}}
\\
\subfloat[$P_{fa}$ versus scaling factor $\rho$]{\label{fig:CFAR_rho}
\includegraphics[width=0.87\linewidth]{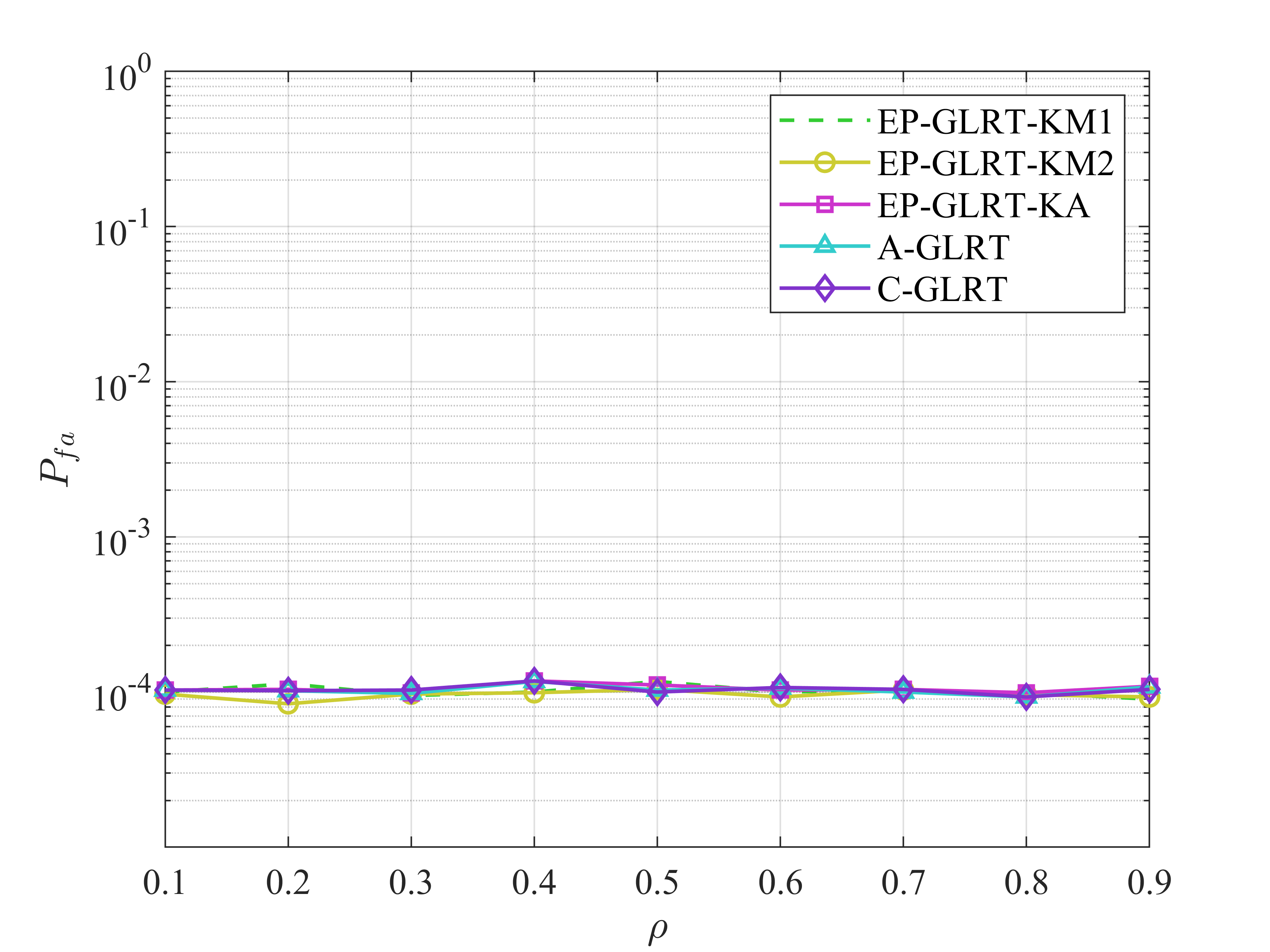}}    
\caption{CFAR analysis of the proposed algorithms assuming $N=16$, $K_S=24$, and
a nominal $P_{fa}=10^{-4}$.}
\label{fig:CFAR}
\end{figure}

In this subsection, we investigate the CFAR behavior 
of the proposed decision schemes with respect to variations of 
both the clutter power (i.e., $\sigma^2_c$) and the clutter 
correlation (i.e., $\rho$). 
To this end, we estimate the actual $P_{fa}$ for the considered 
values of CNR or $\rho$ 
when the used threshold corresponds to a nominal $P_{fa}=10^{-4}$ obtained 
by setting $\mbox{CNR}=25$ dB and $\rho=0.9$. These estimates are computed
with a training sample size $K_S=24$. 
Figure \ref{fig:CFAR_CNR} shows the variation of the estimated $P_{fa}$ 
versus CNR (from $-15$~dB to $30$~dB) when $\rho=0.9$, 
while Figure \ref{fig:CFAR_rho} contains the $P_{fa}$ values 
versus $\rho \in [0.1,0.9]$ for a given CNR equal to $25$~dB. 

The top subfigure confirms the theoretical results obtained in Section 
\ref{Sec:CFARproperty}, namely that all the proposed schemes are 
invariant to the clutter scaling factor 
(at least in a clutter-dominated environment), whereas the bottom 
subfigure highlights the robustness of the detection thresholds
with respect to variations of the covariance structure 
associated with the different values of $\rho$. Recalling that each 
detector is bounded from above by a statistic that is invariant to the covariance structure,
these results show that the bound is tight. 

\subsection{Detection Performance and Estimation Accuracy}
Now, we estimate the $P_d$ as a function 
of the SINR by assuming $K_S=24,32$. 
For comparison purposes, the conventional detectors, i.e., 
Kelly’s GLRT \cite{kelly1986adaptive} and the 
AMF detector \cite{robey1992cfar}, are also included when they 
are fed by the range bins corresponding to either the RTR, the RTSR, or the RSTR path.
Recall that these conventional detectors operate on a single range cell basis and
use the nominal pointing direction of the radar (i.e., $\bv_R$).
Before describing the results, we recall here that the range bins 
$1$, $3$, and $6$ correspond to the path RTR, RTSR (or RSTR), and RSTSR, respectively. 

\begin{figure}[t]
	\centering  
		\includegraphics[height=5.5cm,width=1\linewidth]{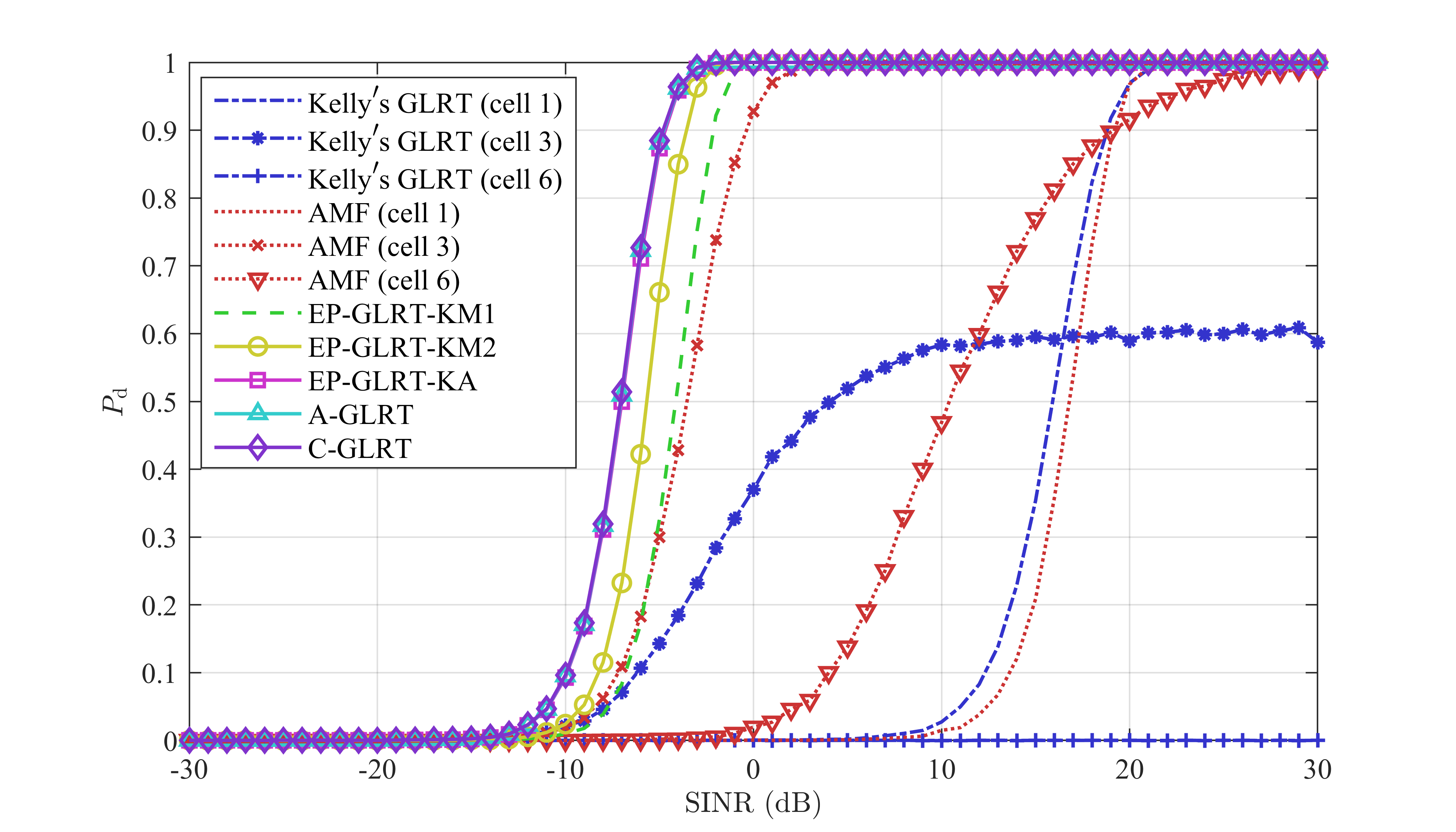}
	\caption{$P_d$ versus SINR assuming $N=16$, $K_S=24$, and $P_{fa}=10^{-4}$.}
\label{fig:Pd 24}
\end{figure}

\begin{figure}[t]
	\centering  
		\includegraphics[height=5.5cm,width=1\linewidth]{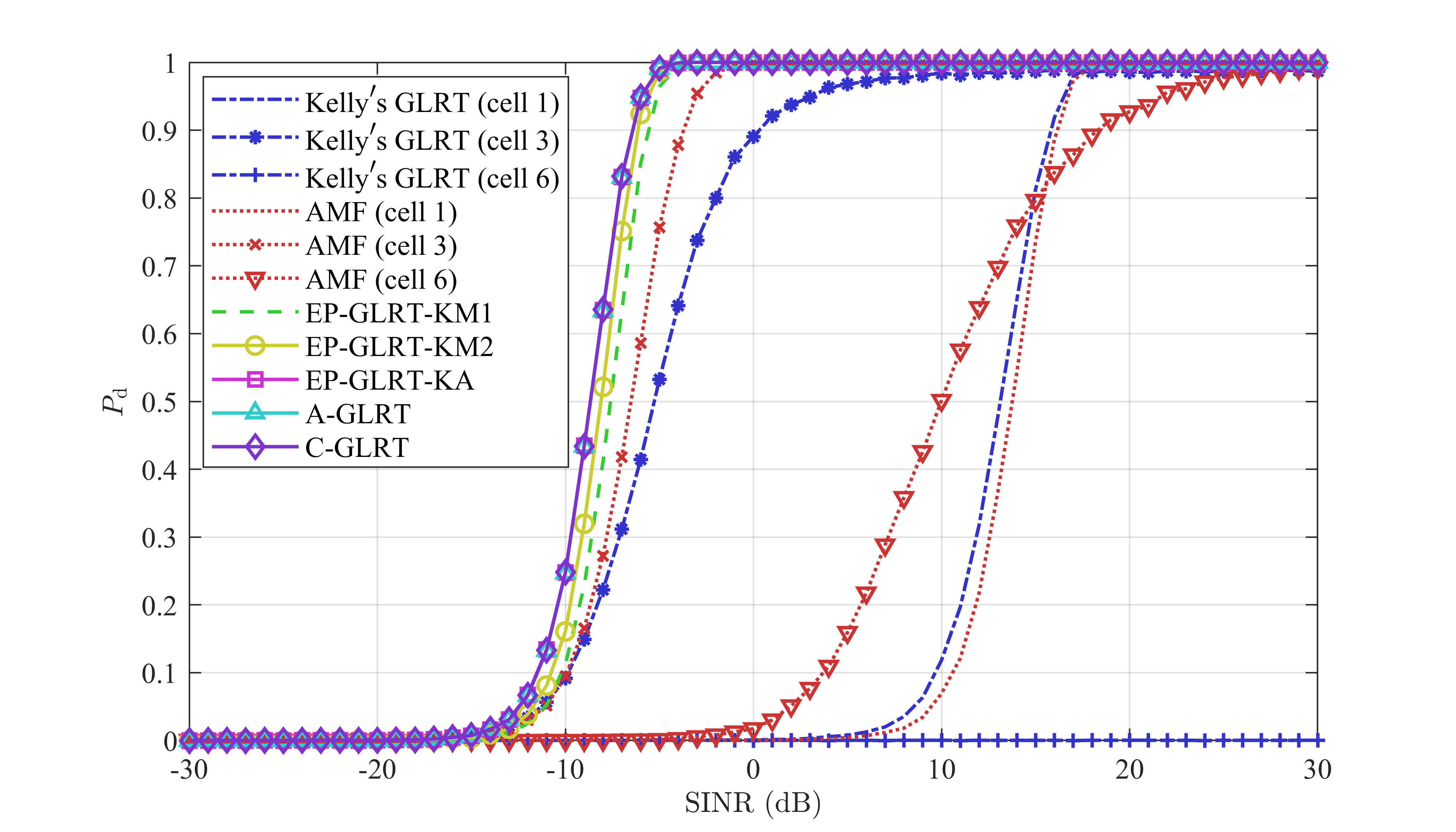}
	\caption{$P_d$ versus SINR assuming $N=16$, $K_S=32$, and $P_{fa}=10^{-4}$.}
\label{fig:Pd 32}
\end{figure}

Figure~\ref{fig:Pd 24} shows the $P_d$ curves for $K_S=24$. 
The proposed RIS-aided schemes can significantly outperform both the conventional 
Kelly's GLRT and the AMF especially when they are fed by single range bins. 
As a matter of fact, Kelly's GLRT and the AMF fed by range cell $1$ 
achieve $P_d>0.9$ at around $\mathrm{SINR}=20$ dB, 
whereas the proposed schemes attain the same detection probability 
at $\mathrm{SINR}=-4$ dB with a performance gain of about $24$ dB. 
The behavior of Kelly's GLRT and AMF changes when they test cells $3$ and $6$.
Specifically, the AMF takes advantage of both its robustness (within certain limits) 
to mismatched signals \cite{BOR-Morgan}
and bistatic echo power to improve its performance. In fact, in range cell $3$,
the actual target signature is $\bv_{SR}$ that has a lower mismatch level than $\bv_S$ 
with respect to the nominal direction $\bv_R$ and the target 
power is significantly higher than that in range cell $1$.
On the contrary, Kelly's GLRT is a selective decision scheme \cite{BOR-Morgan} and high power levels
``make somehow more evident'' the steering mismatch leading to poor detection
performance. Indeed, when fed by range bin $3$, the $P_d$ values
returned by Kelly's GLRT exhibit an upper bound around $0.6$. 
This upper bound also depends on the estimation quality of the disturbance
covariance matrix, namely on the number of training samples.
As for the proposed detectors,
notice that the A-GLRT, C-GLRT, and EP-GLRT-KA share almost the same performance and 
overcome all the other schemes with
a gain of about $5$ dB (at $P_d=0.9$) over the AMF when fed by range cell $3$. The performance of
the EP-GLRT-KM1 and EP-GLRT-KM2 is in between that of A-GLRT and the AMF fed by range cell $3$ with
a gain of about $1.5$ dB for the EP-GLRT-KM2 over the EP-GLRT-KM1.
Two important remarks arise from these results. First, at the design stage, assuming that
$\balpha$ is known and computing the adaptive GLRT with the considered
estimate of $\balpha$
returns better detection performance
than the adaptive GLRT obtained by assuming $\bM$ known and replacing it
with the sample covariance matrix of the secondary and/or primary data. 
Finally, the approximation
used to design the A-GLRT leads to a decision rule that is almost equivalent
to the C-GLRT (for the considered simulation parameters) derived by 
means of an iterative procedure that converges at least to local stationary points.
The impact of the amount of training data on the performance is investigated 
in Figure \ref{fig:Pd 32} where the $P_d$ curves for all the considered
architectures are estimated by assuming $K_S=32$.
As expected, a general increase of the performance occurs with 
the upper bound of Kelly's GLRT when fed by
range cell $3$ that moves from $P_d=0.6$ to $P_d=0.985$
(this increase is due to the large amount of secondary data). 
Moreover, the curves related to the proposed detectors 
are within an interval of about $1$ dB at $P_d=0.9$ with a gain of about
$2.5$ dB with respect to the AMF fed by cell $3$. 
It is also important to point out that, if a detection is declared
in cell $3$ by the AMF, this would 
inherently correspond to a range mispositioning of the considered target.
The gain with respect to
the conventional schemes fed by cell $1$ is still higher than $20$ dB.

\begin{figure}[t]
\centering  
\subfloat[$n$]{
\includegraphics[width=0.85\linewidth]{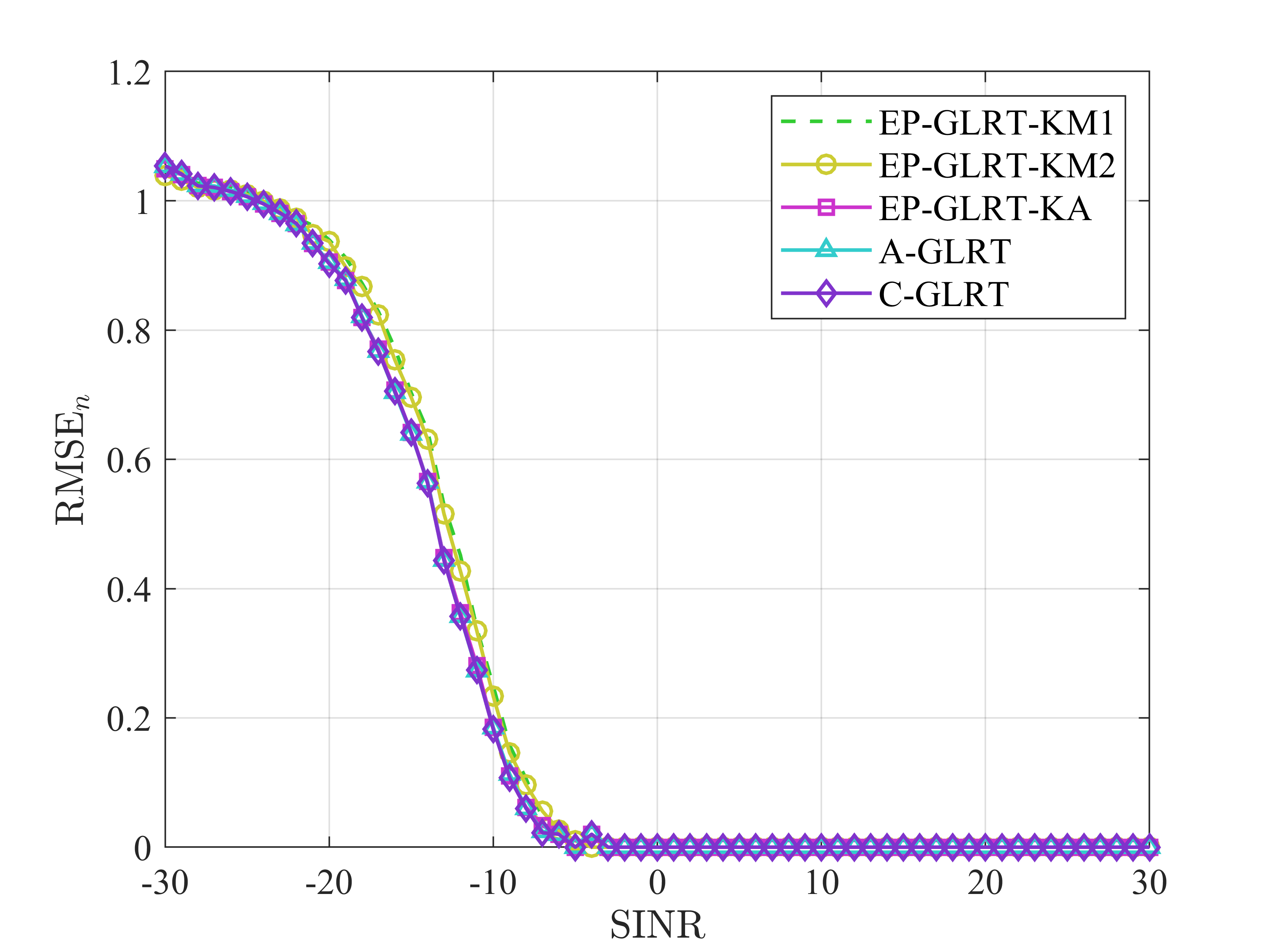}}
\\
\subfloat[$m$]{
\includegraphics[width=0.85\linewidth]{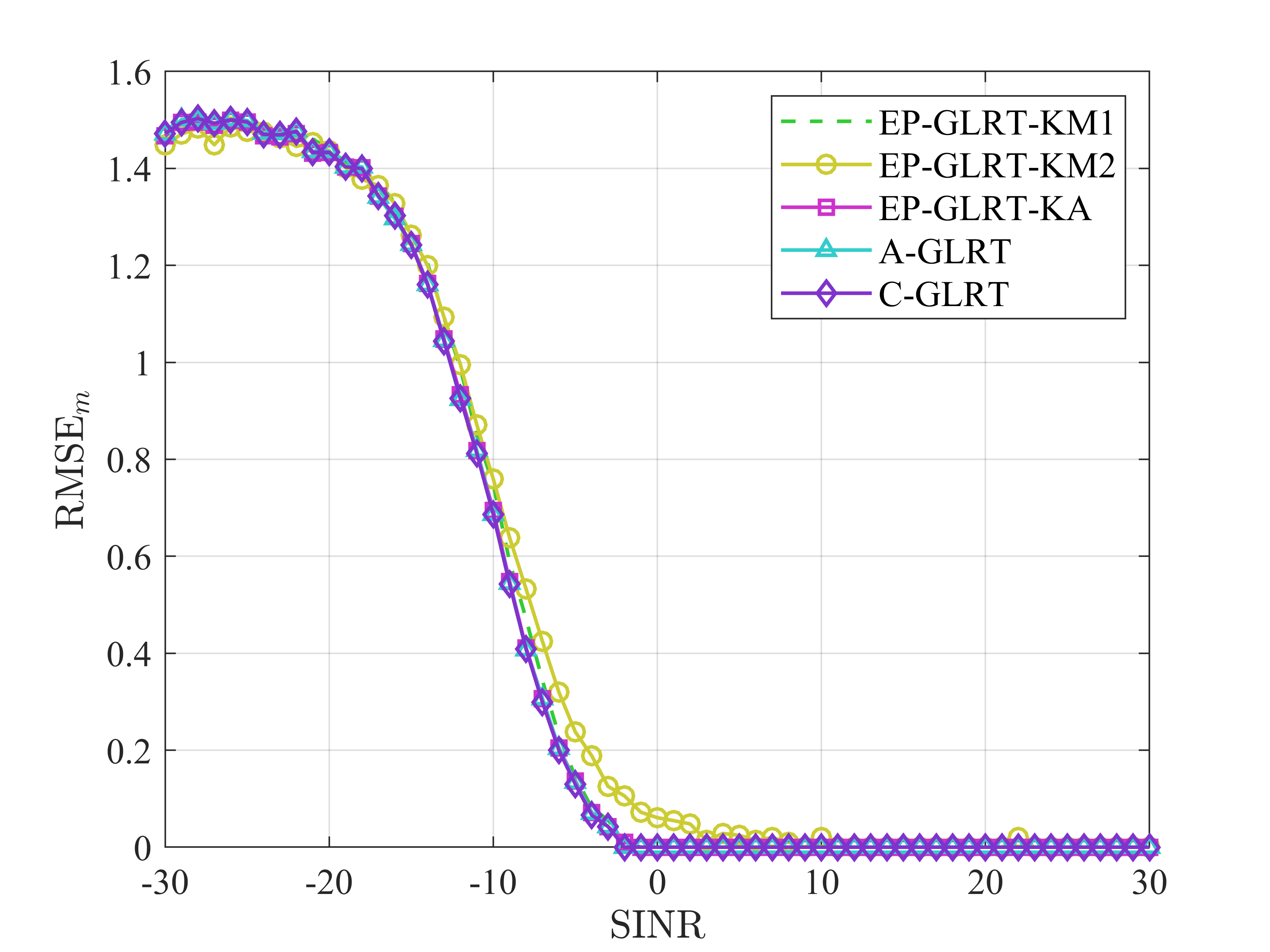}}
\caption{RMSE versus SINR for the estimation of $n$ and $m$ assuming $K_S=32$.}
\label{fig:RMSE_nm 32}
\end{figure}

For the sake of completeness, we evaluate the estimation 
accuracy for the pairs $(n,m)$ assuming $K_S=32$. 
To this end, the performance metrics are the following estimates of the root mean 
square error (RMSE)
\begin{equation}\label{RMSE}
\begin{aligned}
           & \mathrm{RMSE}_{n} = \sqrt{\frac{1}{N_{\mathrm{MC}}} \sum_{k=1}^{N_{\mathrm{MC}}}   (\widehat{n}_k -n)^2},\\
           & \mathrm{RMSE}_{m} = \sqrt{\frac{1}{N_{\mathrm{MC}}} \sum_{k=1}^{N_{\mathrm{MC}}}   (\widehat{m}_k -m)^2},
\end{aligned}
\end{equation}
where $\widehat{n}_k$ and $\widehat{m}_k$ denote the estimates 
of $n$ and $m$ in the $k$th MC trial, respectively, 
and $N_{\mathrm{MC}}=10000$ is the total number of trials. 
As shown in Figure~\ref{fig:RMSE_nm 32}, all the proposed detectors
share almost the same estimation performance with RMSE values
that decrease below $1$ when $\mbox{SINR}>-25$ dB for $n$ and
$\mbox{SINR}>-15$ dB for $m$. Notice that the conventional
detectors are not capable of estimating the positions of the echoes
related to the different paths and, hence, they are not considered in this analysis.

\begin{figure}[t]
\begin{center}
\includegraphics[scale=0.45]{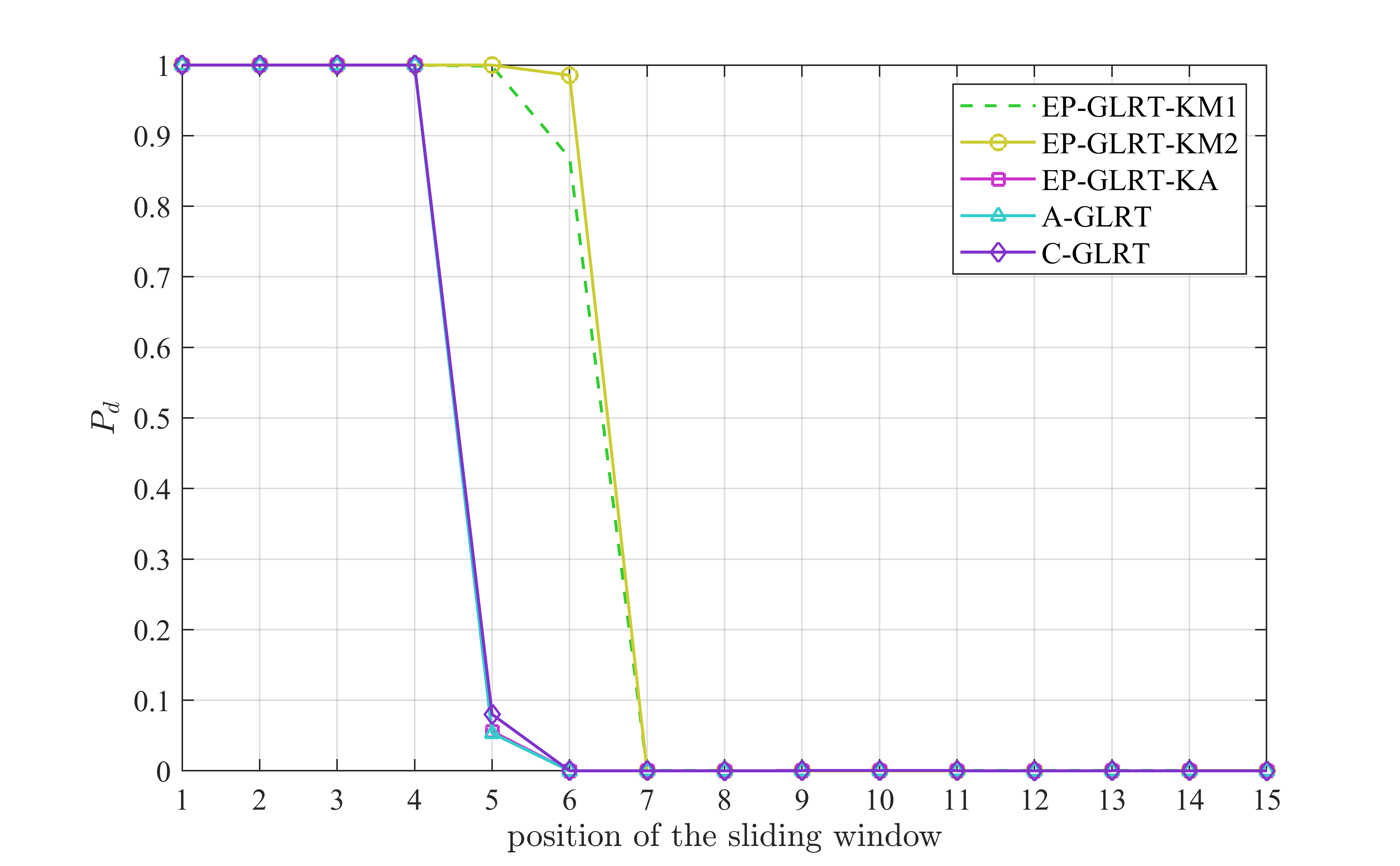}
\caption{$P_d$ versus the position of the sliding window assuming $K_S=6$ and $P_{fa}=10^{-4}$.}
\label{fig:sliding window}
\end{center}
\end{figure}

Finally, we analyze the detection performance within a window of size $K_P=6$
that moves over the $20$ range bins of interest; we assume $K_S=24$, and
$\mbox{SINR}=0$ dB; the last value guarantees
that the proposed detectors have achieved the maximum $P_d$ value as shown
in Figure \ref{fig:Pd 24}. This analysis has a practical value since
it shows the behavior of the proposed architectures from an operating point of view.
The results are reported in Figure \ref{fig:sliding window} where we plot
the $P_d$ values (recall that $P_{fa}=10^{-4}$) against the position of the
first range cell belonging to the sliding window.
From the inspection of the figure, it turns out that the $P_d$ of A-GLRT, 
C-GLRT, and EP-GLRT-KA drops to zero when cells $1$ and $3$ exit from the sliding
window, while for the EP-GLRT-KM1 and EP-GLRT-KM2, the $P_d$ values quickly decrease
to zero when also range bin $6$ no longer belongs to the moving window.
To explain this behavior, let us recall that the alternative hypothesis considers 
the situation where 
the window under test contains three range bins with target components.
Thus, when the window moves, the cells contaminated by the target
exit the window leading to a configuration that does not perfectly
match the $H_1$ hypothesis. This mismatch can be the cause that
induces the A-GLRT, C-GLRT, and EP-GLRT-KA to decide for $H_0$ when
also range bin $3$ exits the window, whereas EP-GLRT-KM1 and EP-GLRT-KM2
are less sensitive to this mismatch. The low sensitivity to this kind of mismatch 
can be a drawback because it forces the detector to declare the presence
of target components in two additional cells occupied by disturbance only.

\bigskip

Summarizing, the above analysis has singled out the A-GLRT, C-GLRT, and EP-GLRT-KA
as the most promising solutions for the detection of dim targets by exploiting
the echoes reflected by a RIS that is suitably designed and deployed within
the region of interest. In fact, they share almost the same performance
in terms of both probability of detection and estimation accuracy overcoming the other
considered decision schemes.

\section{Conclusions}
\label{Sec:Conclusions}
The problem of detecting low-RCS targets has been addressed without
resorting to neither multistatic radar networks, with their synchronization,
costs, and energy consumption issues, nor multi-frame approaches characterized
by high computational costs. To this end, a search radar using a fan beam
has been considered. The radar illuminates a RIS
deployed in the region of interest 
with the aim of intercepting the energy along the irrelevant directions where
the low-RCS target redirects the energy coming from the flight direction.
The RIS has been designed to reflect this energy towards the radar that can 
capitalize most of the energy backscattered by the target. To this end, we have provided
a guideline to size the RIS also suggesting possible weights for its elements.
Then, we have exploited the GLRT design criterion and ad hoc modifications of it
to come up with five detection architectures that can ensure the CFAR property
at least with respect to the clutter power.
The performance analysis has shown the effectiveness of these architectures
in comparison with well-known conventional detectors. Moreover, three out of the five
proposed detectors, namely the A-GLRT, C-GLRT, and EP-GLRT-KA, have arisen as the most
promising solutions.

Future research tracks might encompass the use of multiple RIS deployed around
the radar system to intercept more energy as well as the 
exploitation, at the design stage, of geometry/clutter
symmetries that can reduce the number of unknown parameters to be estimated.

\bibliographystyle{IEEEtran}
\bibliography{group_bib}

\end{document}